\documentclass[twocolumn,prd,superscriptaddress,altaffilletter,nofootinbib]{revtex4-1}

\usepackage{amsmath}
\usepackage{amssymb}
\usepackage{amsfonts}
\usepackage{graphicx}
\usepackage{lineno}
\usepackage[usenames,dvipsnames]{xcolor}
\usepackage{xspace}
\usepackage{tikz}
\usetikzlibrary{arrows,shapes,trees,decorations.pathreplacing}
\usepackage{enumitem}
\usepackage{soul}

\usepackage{caption}
\usepackage{subcaption}
\usepackage[colorlinks,citecolor=blue,urlcolor=red,bookmarks=false,hypertexnames=true]{hyperref} 
\usepackage{dcolumn}
\usepackage{bm}
\usepackage{ulem}
\usepackage{color}
\usepackage{multirow}
\usepackage{mwe}
\usepackage{inputenc}
\usepackage{orcidlink}

\allowdisplaybreaks

\allowdisplaybreaks

\begin{document}

\preprint{APS/123-QED}

\title{Template bank for compact binary mergers in the fourth observing run of Advanced LIGO, Advanced Virgo, and KAGRA}

\author{Shio Sakon \orcidlink{0000-0002-5861-3024}}
\email{shio.sakon@ligo.org}
\affiliation{Department of Physics, The Pennsylvania State University, University Park, PA 16802, USA}
\affiliation{Institute for Gravitation and the Cosmos, The Pennsylvania State University, University Park, PA 16802, USA}

\author{Leo Tsukada \orcidlink{0000-0003-0596-5648}}
\affiliation{Department of Physics, The Pennsylvania State University, University Park, PA 16802, USA}
\affiliation{Institute for Gravitation and the Cosmos, The Pennsylvania State University, University Park, PA 16802, USA}

\author{Heather Fong}
\affiliation{Department of Physics and Astronomy, University of British Columbia, Vancouver, BC, V6T 1Z4, Canada}
\affiliation{RESCEU, The University of Tokyo, Tokyo, 113-0033, Japan}
\affiliation{Graduate School of Science, The University of Tokyo, Tokyo 113-0033, Japan}

\author{James Kennington \orcidlink{0000-0002-6899-3833}}
\affiliation{Department of Physics, The Pennsylvania State University, University Park, PA 16802, USA}
\affiliation{Institute for Gravitation and the Cosmos, The Pennsylvania State University, University Park, PA 16802, USA}

\author{Wanting Niu \orcidlink{0000-0003-1470-532X}}
\affiliation{Department of Physics, The Pennsylvania State University, University Park, PA 16802, USA}
\affiliation{Institute for Gravitation and the Cosmos, The Pennsylvania State University, University Park, PA 16802, USA}

\author{Chad Hanna}
\affiliation{Department of Physics, The Pennsylvania State University, University Park, PA 16802, USA}
\affiliation{Institute for Gravitation and the Cosmos, The Pennsylvania State University, University Park, PA 16802, USA}
\affiliation{Department of Astronomy and Astrophysics, The Pennsylvania State University, University Park, PA 16802, USA}
\affiliation{Institute for Computational and Data Sciences, The Pennsylvania State University, University Park, PA 16802, USA}

\author{Shomik Adhicary}
\affiliation{Department of Physics, The Pennsylvania State University, University Park, PA 16802, USA}
\affiliation{Institute for Gravitation and the Cosmos, The Pennsylvania State University, University Park, PA 16802, USA}

\author{Pratyusava Baral \orcidlink{0000-0001-6308-211X}}
\affiliation{Leonard E.\ Parker Center for Gravitation, Cosmology, and Astrophysics, University of Wisconsin-Milwaukee, Milwaukee, WI 53201, USA}

\author{Amanda Baylor \orcidlink{0000-0003-0918-0864}}
\affiliation{Leonard E.\ Parker Center for Gravitation, Cosmology, and Astrophysics, University of Wisconsin-Milwaukee, Milwaukee, WI 53201, USA}

\author{Kipp Cannon \orcidlink{0000-0003-4068-6572}}
\affiliation{RESCEU, The University of Tokyo, Tokyo, 113-0033, Japan}

\author{Sarah Caudill}
\affiliation{Department of Physics, University of Massachusetts, Dartmouth, MA 02747, USA}
\affiliation{Center for Scientific Computing and Data Science Research, University of Massachusetts, Dartmouth, MA 02747, USA}

\author{Bryce Cousins \orcidlink{0000-0002-7026-1340}}
\affiliation{Department of Physics, University of Illinois, Urbana, IL 61801 USA}
\affiliation{Department of Physics, The Pennsylvania State University, University Park, PA 16802, USA}
\affiliation{Institute for Gravitation and the Cosmos, The Pennsylvania State University, University Park, PA 16802, USA}

\author{Jolien D. E. Creighton \orcidlink{0000-0003-3600-2406}}
\affiliation{Leonard E.\ Parker Center for Gravitation, Cosmology, and Astrophysics, University of Wisconsin-Milwaukee, Milwaukee, WI 53201, USA}

\author{Becca Ewing}
\affiliation{Department of Physics, The Pennsylvania State University, University Park, PA 16802, USA}
\affiliation{Institute for Gravitation and the Cosmos, The Pennsylvania State University, University Park, PA 16802, USA}

\author{Richard N. George \orcidlink{0000-0002-7797-7683}}
\affiliation{Center for Gravitational Physics, University of Texas at Austin, Austin, TX 78712, USA}

\author{Patrick Godwin}
\affiliation{LIGO Laboratory, California Institute of Technology, Pasadena, CA 91125, USA}
\affiliation{Department of Physics, The Pennsylvania State University, University Park, PA 16802, USA}
\affiliation{Institute for Gravitation and the Cosmos, The Pennsylvania State University, University Park, PA 16802, USA}

\author{Reiko Harada}
\affiliation{RESCEU, The University of Tokyo, Tokyo, 113-0033, Japan}
\affiliation{Graduate School of Science, The University of Tokyo, Tokyo 113-0033, Japan}

\author{Yun-Jing Huang \orcidlink{0000-0002-2952-8429}}
\affiliation{Department of Physics, The Pennsylvania State University, University Park, PA 16802, USA}
\affiliation{Institute for Gravitation and the Cosmos, The Pennsylvania State University, University Park, PA 16802, USA}

\author{Rachael Huxford}
\affiliation{Department of Physics, The Pennsylvania State University, University Park, PA 16802, USA}
\affiliation{Institute for Gravitation and the Cosmos, The Pennsylvania State University, University Park, PA 16802, USA}

\author{Prathamesh Joshi \orcidlink{0000-0002-4148-4932}}
\affiliation{Department of Physics, The Pennsylvania State University, University Park, PA 16802, USA}
\affiliation{Institute for Gravitation and the Cosmos, The Pennsylvania State University, University Park, PA 16802, USA}

\author{Soichiro Kuwahara}
\affiliation{RESCEU, The University of Tokyo, Tokyo, 113-0033, Japan}
\affiliation{Graduate School of Science, The University of Tokyo, Tokyo 113-0033, Japan}

\author{Alvin K. Y. Li \orcidlink{0000-0001-6728-6523}}
\affiliation{LIGO Laboratory, California Institute of Technology, Pasadena, CA 91125, USA}

\author{Ryan Magee \orcidlink{0000-0001-9769-531X}}
\affiliation{LIGO Laboratory, California Institute of Technology, Pasadena, CA 91125, USA}

\author{Duncan Meacher \orcidlink{0000-0001-5882-0368}}
\affiliation{Leonard E.\ Parker Center for Gravitation, Cosmology, and Astrophysics, University of Wisconsin-Milwaukee, Milwaukee, WI 53201, USA}

\author{Cody Messick}
\affiliation{MIT Kavli Institute for Astrophysics and Space Research, Massachusetts Institute of Technology, Cambridge, MA 02139, USA}

\author{Soichiro Morisaki \orcidlink{0000-0002-8445-6747}}
\affiliation{Institute for Cosmic Ray Research, The University of Tokyo, 5-1-5 Kashiwanoha, Kashiwa, Chiba 277-8582, Japan}
\affiliation{Leonard E.\ Parker Center for Gravitation, Cosmology, and Astrophysics, University of Wisconsin-Milwaukee, Milwaukee, WI 53201, USA}

\author{Debnandini Mukherjee \orcidlink{0000-0001-7335-9418}}
\affiliation{NASA Marshall Space Flight Center, Huntsville, AL 35811, USA}
\affiliation{Center for Space Plasma and Aeronomic Research, University of Alabama in Huntsville, Huntsville, AL 35899, USA}

\author{Alex Pace}
\affiliation{Department of Physics, The Pennsylvania State University, University Park, PA 16802, USA}
\affiliation{Institute for Gravitation and the Cosmos, The Pennsylvania State University, University Park, PA 16802, USA}

\author{Cort Posnansky}
\affiliation{Department of Physics, The Pennsylvania State University, University Park, PA 16802, USA}
\affiliation{Institute for Gravitation and the Cosmos, The Pennsylvania State University, University Park, PA 16802, USA}

\author{Anarya Ray \orcidlink{0000-0002-7322-4748}}
\affiliation{Leonard E.\ Parker Center for Gravitation, Cosmology, and Astrophysics, University of Wisconsin-Milwaukee, Milwaukee, WI 53201, USA}

\author{Surabhi Sachdev \orcidlink{0000-0002-0525-2317}}
\affiliation{School of Physics, Georgia Institute of Technology, Atlanta, GW 30332, USA}
\affiliation{Leonard E.\ Parker Center for Gravitation, Cosmology, and Astrophysics, University of Wisconsin-Milwaukee, Milwaukee, WI 53201, USA}

\author{Divya Singh \orcidlink{0000-0001-9675-4584}}
\affiliation{Department of Physics, The Pennsylvania State University, University Park, PA 16802, USA}
\affiliation{Institute for Gravitation and the Cosmos, The Pennsylvania State University, University Park, PA 16802, USA}

\author{Ron Tapia}
\affiliation{Department of Physics, The Pennsylvania State University, University Park, PA 16802, USA}
\affiliation{Institute for Computational and Data Sciences, The Pennsylvania State University, University Park, PA 16802, USA}

\author{Takuya Tsutsui \orcidlink{0000-0002-2909-0471}}
\affiliation{RESCEU, The University of Tokyo, Tokyo, 113-0033, Japan}

\author{Koh Ueno \orcidlink{0000-0003-3227-6055}}
\affiliation{RESCEU, The University of Tokyo, Tokyo, 113-0033, Japan}

\author{Aaron Viets \orcidlink{0000-0002-4241-1428}}
\affiliation{Concordia University Wisconsin, Mequon, WI 53097, USA}

\author{Leslie Wade}
\affiliation{Department of Physics, Hayes Hall, Kenyon College, Gambier, Ohio 43022, USA}

\author{Madeline Wade \orcidlink{0000-0002-5703-4469}}
\affiliation{Department of Physics, Hayes Hall, Kenyon College, Gambier, Ohio 43022, USA}

\author{Jonathan Wang}
\affiliation{Department of Physics, University of Michigan, Ann Arbor, Michigan 48109, USA}

\date{\today}

\begin{abstract}
    Matched-filtering gravitational wave search pipelines identify gravitational wave signals by computing correlations, i.e., signal-to-noise ratios, between gravitational wave detector data and gravitational wave template waveforms.  
Intrinsic parameters, the component masses and spins, of the gravitational wave waveforms are often stored in ``template banks", and the construction of a densely populated template bank is essential for some gravitational wave search pipelines.
This paper presents a template bank that is currently being used by the {\fontfamily{qcr}\selectfont GstLAL}-based compact binary search pipeline in the fourth observing run of the LIGO, Virgo, and KAGRA collaboration, and was generated with a new binary tree approach of placing templates, {\fontfamily{qcr}\selectfont manifold}.
The template bank contains \(1.8 \times 10^6\) sets of template parameters covering plausible neutron star and black hole systems up to a total mass of \(400\) \(M_\odot\) with component masses between $1$-$200$ $M_\odot$ and mass ratios between $1$ and $20$ 
under the assumption that each component object's angular momentum is aligned with the orbital angular momentum. 
We validate the template bank generated with our new method, {\fontfamily{qcr}\selectfont manifold}, 
by comparing it with a template bank generated with the previously used stochastic template placement method.
We show that both template banks have similar effectualness.
The {\fontfamily{qcr}\selectfont GstLAL} search pipeline performs singular value decomposition (SVD) on the template banks to reduce the number of filters used. 
We describe a new grouping of waveforms that improves the computational efficiency of SVD by nearly $5$ times as compared to previously reported SVD sorting schemes.
\end{abstract}

\maketitle


\section{Introduction}

Advanced LIGO \cite{Harry2010} detected gravitational waves (GWs) from merging black holes (BHs) for the first time in $2015$ during its first observing run (O1)  ~\cite{TheLIGOScientific:2016pea, TheLIGOScientific:2014jea, LIGOScientific:2018mvr, Abbott:2016blz, gwtc2_1, Abbott:2016nmj, LIGOScientific:2019fpa, LIGOScientific:2018mvr}.
During the second observing run (O2) ~\cite{LIGOScientific:2018mvr}, LIGO and Virgo ~\cite{TheVirgo:2014hva} detected GWs emitted during the inspiral and merger of a binary neutron star (BNS) system, GW$170817$, ~\cite{Abbott:2018hgk, Abbott:2018wiz, TheLIGOScientific:2017qsa}, which marked the start of multi-messenger astrophysics with GW and electromagnetic wave observations ~\cite{TheLIGOScientific:2017qsa, Abbott:2017dke, Monitor:2017mdv, Abbott_2017}.
By the end of the third observation run (O3), a total of \(90\) GW candidate events have been added to the Gravitational Wave Transient Catalog (GWTC) ~\cite{theligoscientificcollaboration2021gwtc3}, including BNS and NS - BH (NSBH) merger events ~\cite{Abbott_2021}.
At the time of writing, the fourth observation run (O4) is ongoing with the LIGO detectors at a higher sensitivity than the previous runs, and have already reported dozens of GW candidates via the LVK public alert system ~\cite{gracedb_public_o4}, while Virgo and KAGRA ~\cite{KAGRAscience} are scheduled to join later in the run ~\cite{LIGO_O4_update}.

Compact binary coalescence (CBC) search pipelines ~\cite{Messick_2017, cannon2021gstlal, Sachdev2019vvd, hanna_2020, gstlal, Usman:2015kfa, PhysRevD.98.024050, Dal_Canton_2021, Adams:2015ulm, Aubin_2021, Andres_2022, PhysRevD.105.024023, PhysRevD.86.024012, Chu_Thesis}
identify GW signals from detector data with matched-filtering algorithms, which is optimized for detecting signals from stationary and Gaussian noise.  
GW waveforms (henceforth, ``templates") are correlated with detector data, yielding signal-to-noise ratios (SNRs), which is a key ingredient for assigning significance of GW events ~\cite{Tsukada_LR_2023, Ewing_performance_2023}. 
Templates are often provided to the matched-filtering algorithm as a discrete set of templates that span a desired parameter space called a ``template bank", which minimizes SNR loss in the targeted parameter space of the bank ~\cite{PhysRevD.44.3819, PhysRevD.49.1707, PhysRevD.53.6749, Owen:1998dk}.
CBC search pipelines have been successfully detecting GWs from CBCs ~\cite{LIGOScientific:2019fpa, abbott2021gwtc, theligoscientificcollaboration2021gwtc3} since O1 using template banks. 

This paper focuses on the development and analysis of the template bank used by the {\fontfamily{qcr}\selectfont GstLAL} search pipeline ~\cite{Messick_2017, cannon2021gstlal, Sachdev2019vvd, hanna_2020, gstlal}. 
{\fontfamily{qcr}\selectfont GstLAL} is a time-domain matched-filtering pipeline that combines standard GStreamer ~\cite{gstreamer} signal processing elements with custom elements to utilize the LIGO Algorithm Library (LAL) ~\cite{lalsuite}, i.e., tools for GW data analysis, and enables parallel processing of GW data ~\cite{cannon2021gstlal}.
{\fontfamily{qcr}\selectfont GstLAL} analyzes GW data in low latency (referred to as ``online") as well as after archival data are available (referred to as ``offline"), detects GW candidates, provides estimates of the source parameters and estimates the significance of the detected candidates ~\cite{Messick_2017, cannon2021gstlal, Sachdev2019vvd, Ewing_performance_2023}. 
Online modes are used to analyze strain data in near real-time during observing runs and enable the pipeline to detect signals in $\sim 10$ seconds ~\cite{Ewing_performance_2023}, such that alerts of GW detections can be sent out to the public ~\cite{Abbott_2019}, opening up the possibility of multi-messenger observations of signals with electromagnetic counterparts. 

The template bank presented in this paper aims to detect BNSs, NSBHs, and BBHs, whose component masses range from $1$-$200$ $M_\odot$ and mass ratios range from $1$-$20$. 
The organization of this paper is as follows: section \ref{section:design} describes the construction, design, motivations for choices of parameter ranges, and the method of generating a template bank for {\fontfamily{qcr}\selectfont GstLAL}'s O4 search, section \ref{section:results} describes the performance of the new SVD sorting parameters and the O4 template bank, and section \ref{section:conclusion} presents our conclusions.

\section{Design and Method}
\label{section:design}

In this section, we provide details of the construction and design of the O4 template bank, an overview of the template bank generation method, describe the checkerboarding method that is implemented on the O4 template bank to run online analysis efficiently across multiple data centers to minimize data processing failures, and SVD processes that are performed on the O4 template bank for filtering efficiency.  
Readers are referred to ~\cite{PhysRevD.44.3819, PhysRevD.49.1707, PhysRevD.53.6749, Owen:1998dk} that provide details on matched-filtering and the foundations of templates and template banks for GW detection, and ~\cite{Sachdev2019vvd, Messick_2017, Mukherjee_2021, cannon2021gstlal} that provide details of matched-filtering in the {\fontfamily{qcr}\selectfont GstLAL} pipeline. 

\subsection{Construction}
\label{subsection:construction}

This section illustrates the construction of a template bank that is densely populated in the search-targeted parameter space such that the {\fontfamily{qcr}\selectfont GstLAL} pipeline can identify GW signals using templates.

Templates
are composed of extrinsic and intrinsic parameters. 
Intrinsic parameters are the component masses and spin of the merging compact objects.
The O4 template bank uses spin-aligned templates, i.e., the angular momenta of the component objects are aligned with the orbital angular momentum of the binary, such that the in-plane spin components are set to zero.
Intrinsic parameters are necessary ~\cite{Ajith_2014} to place templates and changing these values requires recomputing the template, hence, the intrinsic parameters are provided when generating templates.

Extrinsic parameters include parameters such as angles of sky location, distance, and the time and phase of coalescence ~\cite{Harry_2014, Allen_2012}.
For spin-aligned templates at a given coalescence time $t_0$, the extrinsic parameters leave the basic template waveforms unchanged while only affecting the overall phase and amplitude of the detected signals. 
Thus, the extrinsic parameters can be maximized over analytically by Fourier Transforms ~\cite{Owen:1998dk} 
\footnote{Details of extrinsic parameter maximization can be found in Eq. $\left( 2.7\right)$ of ~\cite{Owen:1998dk}. The coalescence time and constant phase shift can be maximized over by redefining SNR as a function of time.}.
For the O4 template bank, the extrinsic parameters manifest themselves as time shifts and phase shifts. 
Utilizing extrinsic parameters contributes to minimizing the computational costs of template generation, as the computation of the templates from the intrinsic parameters and the computation of the effects of the extrinsic parameters can be decoupled.

As the parameters of real GW signals are not known \textit{a priori}, the search-targeted parameter space must be densely populated according to the acceptable difference between the neighboring templates to minimize the loss in SNR.
The overlap between two templates, $u_{k}\left(t\right)$ and $u_{j}\left(t\right)$, where $k$ and $j$ indicate the $k$-th and $j$-th template in the template bank, respectively, is ~\cite{PhysRevD.89.024003}: 
\begin{equation}
    \langle u_k | u_j \rangle = 2 \int_{f_\text{low}}^{\infty} \mathrm{d}f \frac{\tilde{u_{k}}\left(f\right) \tilde{u^{*}_{j}}\left(f\right)+\tilde{u^{*}_{k}}\left(f\right) \tilde{u_{j}}\left(f\right)}{S_n\left(f\right)}
\end{equation}
where $S_n\left(f\right)$ is the detectors' one-sided noise power spectral density (PSD), the tilde indicates the Fourier Transform of the templates, $*$ denotes the complex conjugate, $f_\text{low}$ is the low-frequency cutoff, and the templates are normalized such that $\langle \tilde{u_k} | \tilde{u_j} \rangle = 1$. 
The match, $M\left(u_k, u_j\right)$, between the two templates is computed by maximizing over a set of extrinsic parameters, $\{t_c, \phi_c\}$, of the templates ~\cite{Mukherjee_2021}:
\begin{equation}
    M\left(u_k, u_j\right) = \max_{\{t_c, \phi_c\}} \langle u_k | u_j \rangle .
\end{equation}
where $t_c$ and $\phi_c$ are the template's coalescence time and coalescence phase. 
The mismatch between the two templates is 
\begin{equation}
    \text{mismatch} = 1 - M\left(u_k, u_j\right).
\end{equation}
The worst match would occur when a real GW signal lies in the middle of the surrounding templates in the templates' manifold.  
The minimal match, defined as 
\begin{equation}
    \text{minimal match} \equiv 1 - \text{mismatch},
\end{equation}
determines how densely the templates should be placed, and therefore, dictates the number of templates in a template bank. 
For CBC template banks, the minimal match is typically set to \(97\%\) such that the loss in event rate is at an acceptable amount of \(\sim 10\%\) ~\cite{PhysRevD.53.6749}. 

Templates are correlated with GW data and yield SNRs as a function of time. 
Stretches of data that ring up SNRs above a threshold are called ``triggers".
The templates are maximized over by $\phi$ for a short amount of time around the coalescence time to identify templates that produce large SNRs, and the maximum SNR template is passed to the downstream process to assess the likelihood of the trigger being a GW signal ~\cite{Tsukada_LR_2023, Ewing_performance_2023} \footnote{Readers are referred to ~\cite{Tsukada_LR_2023, Cannon2015gha, hanna_2020, Sachdev2019vvd, Messick_2017, Fong2018thesis} for details on the likelihood ratio raking statistics.}. 
The parameters that are mapped to the template yielding the minimum false alarm rate or maximum SNR provide estimates of the trigger parameters, and further parameter estimations ~\cite{Veitch_2015, Ashton_2019, Romero_Shaw_2020, 10.1093/mnras/staa278, PhysRevD.92.023002} determine the source parameters.
Downstream processes, such as the signal-based \(\xi^2\) tests ~\cite{Messick_2017} to compute the consistency of the SNR time series of the template that produced an above-threshold SNR, calculating $\text{p}_{astro}$ values ~\cite{theligoscientificcollaboration2021gwtc3, Ray_pastro_2023} to obtain the probability that a trigger is caused by a certain source category, and computing the population model ~\cite{Fong2018thesis, Ray_pastro_2023} to provide template probability weights based on astrophysical probabilities, all utilize the template parameters. 
Therefore, constructing a densely populated template bank is crucial for {\fontfamily{qcr}\selectfont GstLAL} to detect and characterize GWs. 

In previous observation runs (O1 ~\cite{Abbott_2016_150914, TheLIGOScientific:2016pea}, O2 and O3 ~\cite{Mukherjee_2021}), template banks were generated using a stochastic template placement method; {\fontfamily{qcr}\selectfont LALApps} ~\cite{lalsuite} {\fontfamily{qcr}\selectfont sbank} ~\cite{Ajith_2014, Capano_2016, PhysRevD.89.024003, Harry_2009}. 
The {\fontfamily{qcr}\selectfont sbank} method is robust but requires repetitive match calculations to determine whether or not to accept a proposed template into the template bank such that the minimal match is satisfied. 
For the O4 template bank generation, we use a computationally efficient method developed by Hanna et. al. ~\cite{MANIFOLD}, called {\fontfamily{qcr}\selectfont manifold}, which places templates in the parameter space via a binary tree approach that depends on the individual template's parameters.
Section \ref{subsection:method} explains the details. 

\subsection{Design}

\begin{figure}
    \centering
    \includegraphics[width=8.6cm]{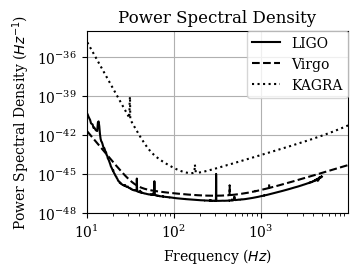}
    \caption{\protect PSD is the power distribution in frequency space that provides a measure for the detectors' sensitivity and dictates the number of templates in a template bank. As the PSD has been updated since the O2/O3 template bank was generated, the LIGO PSD for ``O4 simulation purposes" ~\cite{ligo_psd_dcc} of $190$ Mpc was used to construct and test the O4 template bank. The Virgo PSD was obtained from ~\cite{v1_psd_dcc} and the KAGRA PSD was obtained from ~\cite{k1_psd_dcc}. Among the PSDs listed in ~\cite{psd_dcc}, the high-sensitivity noise curves were chosen to be plotted here.}
    \label{fig:psd}
\end{figure}

\begin{figure*}[htp]
    \centering
    \begin{minipage}{\textwidth}
        \begin{subfigure}[b]{8.6cm}
            \centering
            \includegraphics[width=8.6cm]{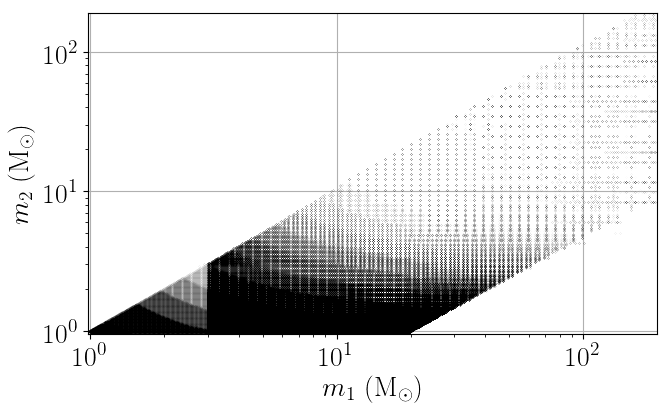}
            \caption{\protect Template placement in the \(m_1\) and \(m_2\) space.}
            \label{fig:template_m1_m2}
        \end{subfigure}
        \hfill
        \begin{subfigure}[b]{8.6cm}
            \centering
            \includegraphics[width=8.6cm]{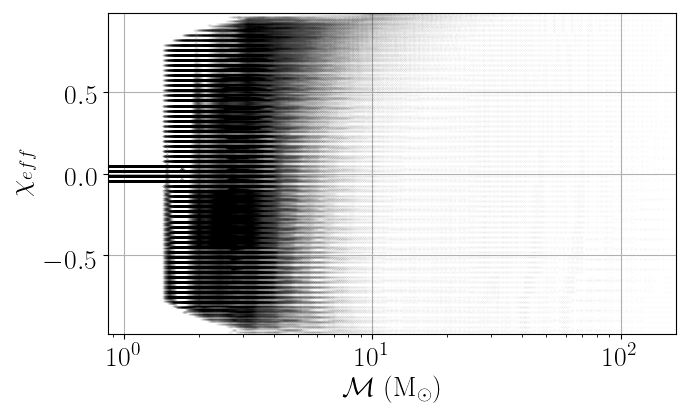}
            \caption{\protect Template placement in the \(\mathcal{M}\) and \(\chi_{\text{eff}}\) space. (See Eq. \protect{\ref{eq:mchirp}} and Eq. \protect{\ref{eq:chi_eff}}.)}
            \label{fig:template_mchirp_chi}
        \end{subfigure}   
            \caption{\protect Template placement of the O4 template bank.}
            \label{fig:template_placement}
    \end{minipage}
\end{figure*}

This paragraph describes the design of the O4 template bank generated with {\fontfamily{qcr}\selectfont manifold} while using the updated O4 PSD \footnote{~\cite{psd_dcc} lists several O4 PSD for simulation purposes, including LIGO PSD with lower sensitivities of $160$ Mpc. The higher sensitivity of the detectors means that two templates can be distinguished better, hence requiring more templates to cover the search parameter space. Refer to ~\cite{Moore_2014, Mukherjee_2021} for details on PSD.} shown in Fig. \ref{fig:psd}. 
Fig. \ref{fig:template_placement} shows the placement of the templates of the O4 template bank, which stores a total of \(\sim 1.8 \times 10^6\) templates.
The intrinsic parameter ranges chosen for the O4 template bank are motivated by observations and the GW detectors' sensitivity ~\cite{Abbott_2020, Legred_2021, Dietrich_2020, Jiang_2020, Abbott_2018, Cromartie_2019, Kalogera_1996, Kramer_2009, Burgay_2003, Stovall_2018, Gou_2011, McClintock_2011, Fabian_2012, theligoscientificcollaboration2021gwtc3}, and are described in detail in the following paragraphs. 

First, the component masses range from \(1\) to \(200\)\(M_\odot\) for both components in the binary such that the template bank covers the parameter space in which we expect CBC events to occur within the detectors' sensitivity range. 
Searches focusing on regions outside this mass range are conducted using different template banks. 
Based on studies on the maximum mass of the NS population that place the upper bounds of NS maximum masses to be between \(2\) to \(3\)\(M_\odot\) ~\cite{Abbott_2020, Legred_2021, Dietrich_2020, Jiang_2020, Abbott_2018, Cromartie_2019, Kalogera_1996}, any merging compact object with a mass below \(3\) \(M_\odot\) are classified as a NS in the O4 template bank. 
This is not to claim that all objects with masses below \(3\) \(M_\odot\) detected by GW searches using {\fontfamily{qcr}\selectfont GstLAL} are NSs.

Secondly, the lower and upper mass gaps are not set when generating the template bank. 
Despite the potential evidence of lower/higher mass gaps ~\cite{Farr_2011, Ozel_2010, Farmer_2019, Heger_2003}, we adopt an agnostic template placement, hence the mass gaps are populated in the same manner as non-mass gap parameter spaces. 

Thirdly, the dimensionless spins aligned with the orbital angular momentum vector of the binary, \(s_{i, z}\) $\left(i = 1, 2\right)$ ~\cite{PhysRevLett.106.241101}, used in the template bank are restricted to \(\pm 0.05\) for components with masses of \(1\) to \(3\)\(M_\odot\), whereas to \(\pm 0.99\) for components with masses \(3\) to \(200\)\(M_\odot\). 
Here, \(s_{1, z}\) and \(s_{2, z}\) are spins of the heavier component in the binary and the lighter component in the binary, respectively, and span \(-1 \leq s_{i, z} \leq 1\).
The choice of spins for NSs is motivated by astrophysical studies ~\cite{Kramer_2009, Burgay_2003, Stovall_2018} that show that the dimensionless spins of NSs in binaries that merge within a Hubble time are observed to be at most \(\pm0.05\).
BHs in binaries are shown to have extremely large spins ~\cite{Gou_2011, McClintock_2011, Fabian_2012}, and the theoretical upper bound for spins of Kerr BHs would be $1$, which leads to our choice of BH spins.
Precession is not included in the O4 template bank, nor higher-order modes, i.e., only the dominant $\left(2, 2\right)$ mode of GW emission is considered. 
Although real GW signals do contain higher-order modes and precession ~\cite{theligoscientificcollaboration2021gwtc3}, the O4 template bank does not include these effects in the templates.

Fourthly, the mass ratio (\(m_1 / m_2\)) of the O4 template bank ranges from \(1\) to \(20\), where $m_1$ is the mass of the heavier component in the binary and $m_2$ is the mass of the lighter component in the binary.
The choice of the maximum mass ratio is motivated by the calibration range of the waveform models ~\cite{Ossokine_2020} and by GW detections in O3 where the bulk of the signals had mass ratios below \(10\) ~\cite{theligoscientificcollaboration2021gwtc3}. 
Setting the maximum mass ratio to \(20\) enables searches to detect GW from sources with large mass ratios that are not yet detected while limiting the number of templates in the template bank.

Finally, the waveform approximant implemented in the method used to place templates is IMRPhenomD.
IMRPhenomD is a computationally efficient phenomenological waveform model that combines analytic post-Newtonian (PN) and effective-one-body methods for the inspiral portion of the waveforms, and the numerical relativity simulations for the merger and ringdown portion of the waveform of spin-aligned BBHs in the frequency-domain ~\cite{Khan_2016, Husa_2016}.

\subsection{Method}
\label{subsection:method}

In this section, we describe the O4 template bank generation method, {\fontfamily{qcr}\selectfont manifold}, and discuss the advantages of using {\fontfamily{qcr}\selectfont manifold} as compared to  {\fontfamily{qcr}\selectfont sbank} to generate template banks.
The equations in this section are for each individual template; hence the subscripts $k$ and $j$ are dropped in the equations. 
Refer to ~\cite{MANIFOLD} for details on the {\fontfamily{qcr}\selectfont manifold} method. 

{\fontfamily{qcr}\selectfont manifold} uses a geometric method to place templates in the intrinsic parameter space. 
The O4 template bank adopts a three-dimensional parameter space of \(\log \left(m_1\right)\), \(\log \left(m_2\right)\), and \(\chi_{\text{eff}}\), where $\chi_{\text{eff}}$ is: 
\begin{equation}
    \chi_{\text{eff}} = \frac{m_1 \times s_{1, z} + m_2 \times s_{2, z}}{m_1 + m_2} .
    \label{eq:chi_eff}
\end{equation}
The targeted parameter space, a hyperrectangle, is split into non-overlapping hyperrectangles via a binary tree approach.
The splitting is done along the hyperrectangle's longest edge according to the template overlap metric, which provides a measure of distance between the nearby templates.
The two new hyperrectangles generated from the bigger hyperrectangle are considered as a pair that lie side-by-side in the intrinsic parameter space and have similar volumes in the $\Delta \log_{10} \left(m_1\right) \times \Delta \log_{10} \left(m_2\right) \times \Delta \chi_{\text{eff}}$ space. 
Here $\Delta \log_{10} \left(m_1\right)$, $\Delta \log_{10} \left(m_2\right)$, and $\Delta \chi_{\text{eff}}$ are the lengths of the sides of the hyperrectangle.
The process of splitting involves numerical computation of the expected number of templates required to cover the split hyperrectangle. 
Splitting of the hyperrectangle terminates when the expected number of templates required to cover the hyperrectangle is below \(1\) according to the minimal match criteria. 
Then, a template is computed and placed using the parameters at the center of the hyperrectangle. 
We also set an upper bound on the volume of the templates: 
\begin{equation}
    \Delta \log_{10} \left(m_1\right) \times \Delta \log_{10} \left(m_2\right) \times \Delta \chi_{\text{eff}} \leq 0.0001
    \label{eq:manifold_coord_requirement}
\end{equation}
to set a lower bound on the template density in the BBH parameter space.
Eq. ~\ref{eq:manifold_coord_requirement} will result in overpopulating the BBH region, as seen in section ~\ref{subsection:bank_sim_results}. 
Extra templates in the BBH region, which otherwise would be sparsely populated, will benefit search tuning, such as when binning templates into SVD bins to collect bin-dependent noise properties ~\cite{Mukherjee_2021}.
The additional computational cost related to placing more templates is minimal as the BBH space is not template-dense. 

The advantages of generating template banks with {\fontfamily{qcr}\selectfont manifold} are the following:
First, generation of template banks with {\fontfamily{qcr}\selectfont manifold} takes \(\mathcal{O}\left(10\right)\) minutes when jobs are run in parallel as compared to weeks when generating template banks with {\fontfamily{qcr}\selectfont sbank} for the same parameter space. 
The computational efficiency is due to the limited number of match calculations that {\fontfamily{qcr}\selectfont manifold} requires when placing templates \footnote{{\fontfamily{qcr}\selectfont manifold} requires match calculations when splitting the hyperrectangles, and needs \(\mathcal{O} \left(n^2\right)\) match calculations per template at worst, where \(n\) is the number of dimensions. See ~\cite{MANIFOLD} for details.} as compared to {\fontfamily{qcr}\selectfont sbank} that calculates the match repetitively for proposed templates with templates in the template bank. 
Secondly, {\fontfamily{qcr}\selectfont manifold} computes the templates' volumes as it places templates, and the nearby templates have similar volumes. 
Meanwhile, {\fontfamily{qcr}\selectfont sbank} does not compute templates' volumes when placing them nor guarantee that their volumes are similar across nearby templates. 
The downstream calculations that use template volumes, such as the population models that account for the astrophysical probabilities of each template ~\cite{Fong2018thesis}, benefit computationally from {\fontfamily{qcr}\selectfont manifold} template banks as template volumes are already computed when the templates are generated. 
Thirdly, {\fontfamily{qcr}\selectfont manifold} generates templates by splitting hyperrectangles into two, so all templates will be generated as pairs that have similar intrinsic parameters and template volumes. 
Having a template pair benefits the process of checkerboarding, which is used for a multi-data center high availability analysis deployment and is explained in section ~\ref{subsection:checker_boarding}.  
Finally, the effectiveness of the template banks generated with {\fontfamily{qcr}\selectfont manifold} is comparable to that of the template banks generated with {\fontfamily{qcr}\selectfont sbank}, as discussed in detail in section \ref{subsection:bank_sim_results}. 

We seek to reduce memory usage during the filtering process by setting an upper limit on the template time duration.
The lower mass templates contain long-duration, low-frequency waveforms that are less crucial to the GW search compared with the high frequencies near the merger. 
To vacate the computing memory in these regions, we set a maximum time duration on the waveform generation process and discard the low-frequency regions that exceed this maximum, leaving the high-frequency regions unaltered. 
For O4, this upper limit of waveform duration is set to $128$ seconds, i.e., if a given low-frequency cutoff \footnote{Here, low-(high-)frequency cutoff are the lower (upper) bounds of the waveforms. For spin-aligned templates, the assumption is that waveforms' frequencies increase monotonically with time. Setting a low-(high-)frequency cutoff is equivalent to setting the start and end of the waveform in time-domain.} results in waveforms longer than $128$ seconds, a higher low-frequency cutoff corresponding to the $128$ second time duration limit will be adopted.  
The waveform generator from {\fontfamily{qcr}\selectfont lalsimulation} within {\fontfamily{qcr}\selectfont lalsuite} ~\cite{lalsuite} adjusts the waveforms' time duration such that the waveforms start from $0$ amplitude.

\subsection{Checkerboarding}
\label{subsection:checker_boarding}

\begin{figure}
    \centering
    \includegraphics[width=8.6cm]{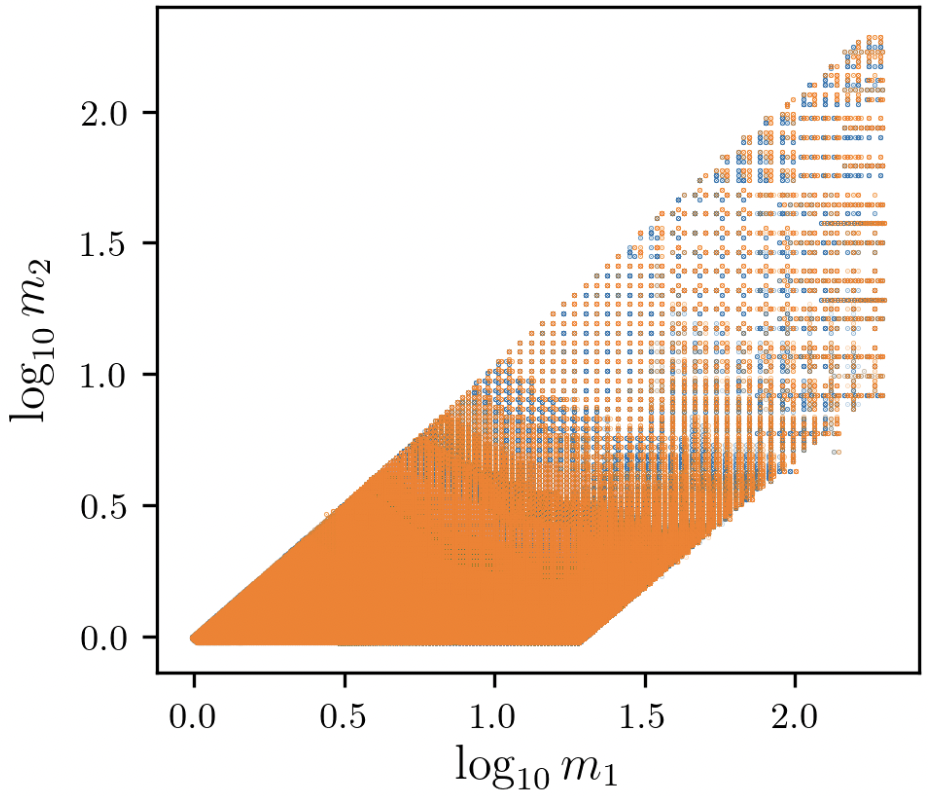}
    \caption{\protect The plot shows the two (blue and orange) checkerboarded template banks overlapped on top of one another in the $\log_{10}\left(m_1\right)$ vs $\log_{10}\left(m_2\right)$ space. The two checkerboarded template banks contain half of the O4 template bank, and the templates in the two checkerboarded template banks are nearly identical sets as a result of {\fontfamily{qcr}\selectfont manifold}'s binary tree approach of template placement. The two checkerboarded template banks are deployed on analyses running on two separate computing clusters. Each checkerboarded template bank consists of $\sim 1 \times 10^6$ templates.}
    \label{fig:checker_boarding}
\end{figure}

Across the entire O4 template bank's parameter space, {\fontfamily{qcr}\selectfont manifold}'s checkerboarding process identifies pairs of templates that constitute the hyperrectangle templates immediately before the splitting was terminated during the template generation and sorts the right side of the pairs into one template bank and the left side of the pairs into another template bank to obtain two nearly identical and complimenting template banks with similar abilities to detect GW signals. 
The two checkerboarded template banks, plotted in Fig. ~\ref{fig:checker_boarding}, are each deployed in online analyses run on two separate computing clusters, and under optimal situations, both analyses will be running such that the {\fontfamily{qcr}\selectfont GstLAL} pipeline has the coverage of the whole O4 template bank at the designed minimal match.
Since the checkerboarded template banks contain half the number of templates compared to the original template bank, two online analyses are performed without doubling the computational cost. 
In events of cluster issues that prevent the analysis on that cluster from processing data, e.g., maintenance and power outages, if the other analysis on a separate cluster is processing data, the {\fontfamily{qcr}\selectfont GstLAL} pipeline will continue to have sensitivity to the entire O4 template bank parameter space but at a $1$ $\%$ decreased minimal match compared to when both analyses are running. 
See section \ref{subsubsection:banksim4} for the checkerboarded template bank effectuality test results. 

\subsection{SVD}
\label{subsection:svd}

While each template in the template bank is unique and necessary to ensure a minimal match, many are redundant for filtering.
SVD is performed on the template bank to construct a reduced set of orthonormal filters, which reduces computation as compared to using the physical templates. 
The following provides a coarse overview of how SVD is done on the template banks and describes the template sorting scheme that has been newly implemented to increase the efficiency of SVD.
Details on SVD are provided in ~\cite{Cannon_2010, Cannon_2012, Messick_2017}. 

\begin{figure}
    \centering
    \includegraphics[width=8.6cm]{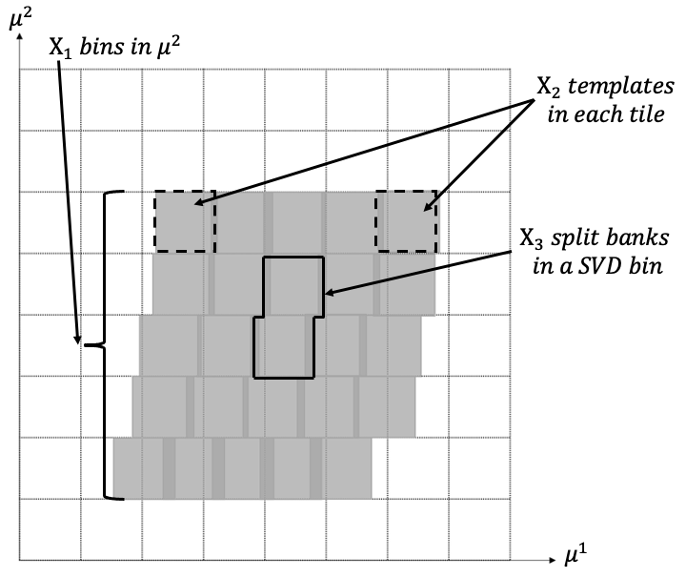}
    \caption{\protect An illustration of how the template bank is bundled into split banks by the sorting parameters, \(\mu^1\) and \(\mu^2\). 
    Each split bank contains \(X_2\) templates and each background bin contains \(X_3\) split bank. 
    SVD is performed on the individual split banks. 
    The overlapping regions of the split banks mitigate the boundary effects.
    } 
    \label{fig:svd}
\end{figure}

Following the template bank generation, the templates are bundled into split banks based on sorting parameters, as illustrated in Fig. \ref{fig:svd}. 
It is necessary to group templates that have similar responses to noise in the same split bank such that the number of basis vectors obtained as a result of the SVD is considerably smaller than the number of original templates in the split bank.
The split banks of the O2/O3 template bank sub-banks were sorted by chirp mass  \(\mathcal{M}\) and \(\chi_\mathrm{eff}\), or template duration ~\cite{Mukherjee_2021}, where  \(\mathcal{M}\) is: 
\begin{equation}
    \mathcal{M} = \frac{\left(m_1 m_2\right)^{3/5}}{\left(m_1 + m_2\right)^{1/5}}. 
    \label{eq:mchirp}
\end{equation}
For the O4 template bank, we introduce a new sorting scheme considering the PN expansion of a waveform. 
For optimizing parameter inference using the Reduced Order Quadrature (ROQ) technique ~\cite{Canizares:2014fya, Smith:2016qas}, ~\cite{Morisaki_2020} constructed ROQ basis sets of waveforms in the targeted mass-spin regions. 
The authors found that the highly compressed ROQ basis sets can be constructed by grouping waveforms based on two principal components of the PN phase coefficients, denoted by $\mu^1$ and $\mu^2$. 
A sorting scheme based on their values efficiently groups template waveforms with similar morphology, reducing the number of SVD filters.
One can find such a combination from the $n$-PN phase terms, which read
\begin{align}
    \begin{split}
        &\Phi^{(n \mathrm{PN})}(f)= \\
        &\sum_{\substack{k=0, \neq 5,8}}^{2 n} \psi^{k}\left(\frac{f}{f_{\mathrm{ref}}}\right)^{\frac{k-5}{3}}+\sum_{k=0}^{2 n} \psi_{\log }^{k}\left(\frac{f}{f_{\mathrm{ref}}}\right)^{\frac{k-5}{3}} \log \left(\frac{f}{f_{\mathrm{ref}}}\right) \\
        &+\psi^{5}+\psi^{8}\left(\frac{f}{f_{\mathrm{ref}}}\right),
    \end{split}
\end{align}
where $f_\mathrm{ref}$ is a reference frequency and, following ~\cite{Morisaki_2020}, we adopt $f_\mathrm{ref}=200 \mathrm{Hz}$. 
The coefficient for each term, e.g. $\psi^k$ and $\psi_\mathrm{log}^k$, corresponds to a combination of parameters on which a waveform has an increasing dependency. 
Up to the 1.5PN order, these coefficients are explicitly given by
\begin{align}
&\psi^{0}(\mathcal{M})=\frac{3}{4}\left(8 \pi \mathcal{M} f_{\mathrm{ref}}\right)^{-\frac{5}{3}}, \\
&\psi^{2}(\mathcal{M}, \eta)=\frac{20}{9}\left(\frac{743}{336}+\frac{11}{4} \eta\right) \eta^{-\frac{2}{5}}\left(\pi \mathcal{M} f_{\mathrm{ref}}\right)^{\frac{2}{3}} \psi^{0}, \\
&\psi^{3}\left(\mathcal{M}, \eta, s_{1, z}, s_{2, z}\right)=(4 \beta-16 \pi) \eta^{-\frac{3}{5}}\left(\pi \mathcal{M} f_{\mathrm{ref}}\right) \psi^{0},
\end{align}
where
\begin{align}
\eta &=\frac{m_{1} m_{2}}{\left(m_{1}+m_{2}\right)^{2}}, \\
\beta &=\frac{1}{12} \sum_{k=1}^{2}\left[113\left(\frac{m_{k}}{M}\right)^{2}+75 \eta\right] s_{i, z}, 
\end{align}
and \(M \equiv m_1 + m_2\).
Furthermore, new parameters orthogonalized using the aforementioned parameters, $\psi^0,\psi^2,\psi^3$, can optimize the sorting efficiency. 
We follow the Fisher analysis to perform the orthogonalization based on ~\cite{Morisaki_2020}, where the authors used the representative PSD of the LIGO-Livingston detector in the O2 and obtained the following linear combinations. 
\begin{align}
\mu^{1}&=0.974 \psi^{0}+0.209 \psi^{2}+0.0840 \psi^{3} \\
\mu^{2}&=-0.221 \psi^{0}+0.823 \psi^{2}+0.524 \psi^{3}.
\end{align}
We adopt these two parameters to sort templates such that the entire template bank is first split into $X_1$ bins by $\mu^2$, then the templates in each $\mu^2$-bin are sorted by $\mu^1$ and grouped every $X_2$ templates to form a split bank.
Each background bin contains $X_3$ split banks that each contain $X_2$ templates, and are used for background estimations. 
The O4 analysis used $X_1 = 20$, $X_2 = 500$, and $X_3 = 2$, such that each SVD bin contains $\sim 1000$ templates. 
The result of using the $\mu^1$ $\mu^2$ sorting, henceforth $\vec{\mu}$ sorting, is discussed in \ref{subsection:svd_results}.
In addition to the above, the O4 analysis adopted an SVD reconstruction tolerance, i.e., the match between the original templates and the reconstructed templates ~\cite{Cannon_2010, Cannon_2012}, of $0.99999$ for higher accuracy of SVD reconstructions. 
 
\section{Results}
\label{section:results}

\subsection{SVD efficiency}
\label{subsection:svd_results}

\begin{figure}
    \centering
    \includegraphics[width=8.6cm]{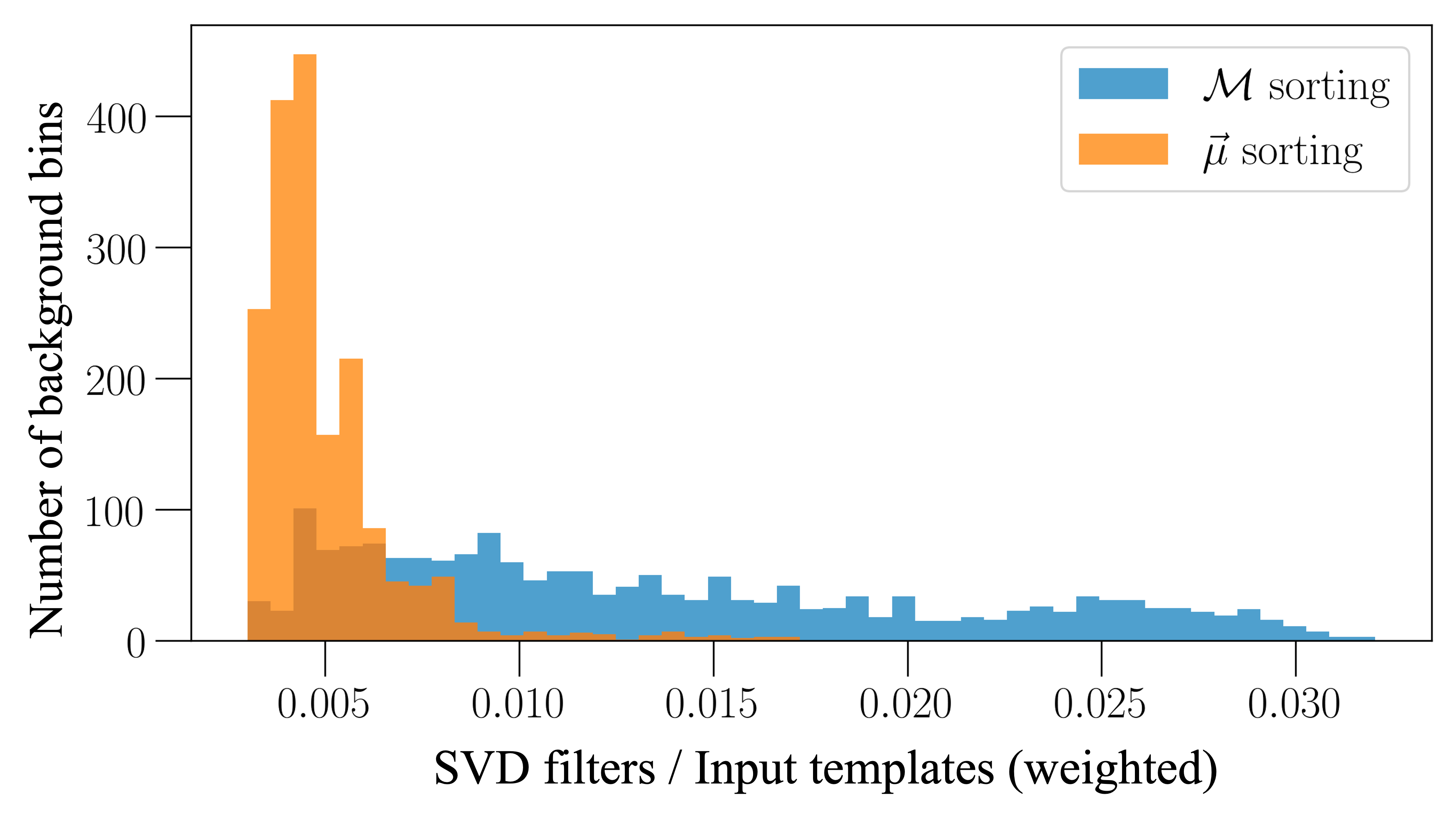}
    \caption{\protect Comparison of SVD efficiency of \(\vec{\mu}\) sorting and previously used \(\mathcal{M}\) sorting.  
    The \(x\) axis is the number of SVD filters divided by the number of templates in the background bin where the numerator and denominator are weighted by the sampling rate, which gives the SVD efficiency. Lower values correspond to higher efficiencies. The \(y\) axis is the number of background bins with a certain efficiency. $\vec{\mu}$ sorting has less variation and higher SVD efficiency as compared to $\mathcal{M}$ sorting.
    } 
    \label{fig:floppulator}
\end{figure}

As described in section \ref{subsection:svd}, $\vec{\mu}$ parameters were used on the O4 template bank to sort the templates into background bins. 
Up until O3, the templates were sorted using $\mathcal{M}$ and $\chi_{\text{eff}}$ or the template duration. 
Authors compared the computational efficiency of the two SVD sorting methods, the $\vec{\mu}$ parameter sorting and the previously used $\mathcal{M}$ sorting, by dividing the number of filters obtained by SVD sorting by the number of templates per background bin where the numerator and denominator are both weighted by the sampling rate. 
See ~\cite{Cannon_2010, PhysRevD.87.122002, Messick_2017} for details on sampling rates. 
As shown in Fig. \ref{fig:floppulator}, $\vec{\mu}$ sorting resulted in higher SVD efficiency by nearly $5$ times with a less variation in the SVD efficiency compared to the previously used $\mathcal{M}$ sorting.

Although the SVD sorting parameters, $\vec{\mu}$, were motivated for lower mass templates and may not be optimal for higher mass templates, we have found that the SVD sorting using $\vec{\mu}$ parameters give more efficient compression rates than using $\mathcal{M}$ and $\chi_{\text{eff}}$. 
Using one sorting scheme throughout the template bank rather than multiple benefits from an operational perspective, as bank-splitting and construction of SVD bins are straightforward.  
Hence, the $\vec{\mu}$ sorting parameters are used for the entire template bank. 
There is potential for future work to improve the SVD sorting scheme.  

\subsection{Bank simulation}
\label{subsection:bank_sim_results}

Bank simulation is a method of computing the fitting factors between the templates in the template bank and the simulated GW waveforms, known as ``injections", to assess the bank's response to GW signals when the bank is deployed in the pipeline. 
Here, the fitting factor is ~\cite{Mukherjee_2021}:
\begin{equation}
    FF\left(u_s\right) = \max_{u \in \{u_k\}} M\left(u, u_s\right)
    \label{eq:ff}
\end{equation}
where \(u_s\) is the waveform of an injection at a given time and Eq. \ref{eq:ff} maximizes over the templates in the template bank, $\{u_k\}$, to obtain the highest matching template with the injection. 
Hereafter, the mismatch is $1 - FF$. 
This section presents the bank simulation results for a template bank generated with {\fontfamily{qcr}\selectfont manifold} (henceforth, the O4 template bank) in comparison to a template bank generated with {\fontfamily{qcr}\selectfont sbank} (henceforth, the {\fontfamily{qcr}\selectfont sbank} template bank) for a similar parameter space to test the effectualness of the O4 template bank.
The {\fontfamily{qcr}\selectfont sbank} template bank consists of \(1.3 \times 10^6\) templates. 
The template banks used the same PSD, same waveform approximant, and same template bank parameter ranges.
The O4 template bank was generated with a lower cutoff of $10$ $\mathrm{Hz}$ and a maximum template duration cut of $128$ seconds, {\fontfamily{qcr}\selectfont sbank} template bank was generated with a lower cutoff of $15$ $\mathrm{Hz}$, and the injections had a lower cutoff of $15$ $\mathrm{Hz}$.
The injections used for the tests presented in this section span the same parameter space, same waveform approximant, and same PSD as for generating template banks.
We performed two bank simulation tests on the template banks as the following:
\begin{enumerate} 
    \item Assess the effectualness of the O4 template bank by comparing the bank simulation results of the O4 template bank and that of the {\fontfamily{qcr}\selectfont sbank} template bank using injections that span the template bank parameter space. (Section \ref{subsubsection:banksim1})
    \item Assess the performance of the two checkerboarded halves of the O4 template bank using injections that span the template bank parameter space. (Section \ref{subsubsection:banksim4}) 
\end{enumerate}
See Appendix \ref{section:appendix} for bank simulation results using injections that lie outside of the template banks.  

\subsubsection{Bank simulation with template bank parameter space injections}
\label{subsubsection:banksim1}

\begin{figure}
    \centering
    \includegraphics[width=8.6cm]{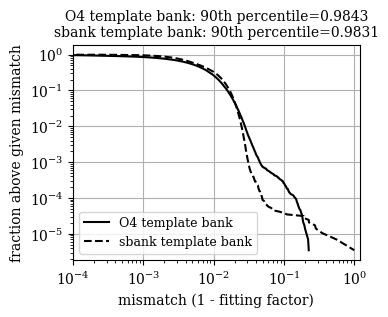}
    \caption{\protect Cumulative histogram of the injections' mismatches for the O4 template bank and the {\fontfamily{qcr}\selectfont sbank} template bank. The \(y\)-axis is the fraction of injections above a given mismatch value. Here, the results of the BNS, NSBH, and BBH injections for each template bank are combined to present the summary. The plot has been truncated at a mismatch of $10^{-4}$ to visually present the results. 
    } 
    \label{fig:cumulative}
\end{figure}

The injections used in this bank simulation are contained in three different injection sets; one injection set each for the BNS, the NSBH, and the BBH parameter space.
The numbers of injections used for the O4 template bank and the {\fontfamily{qcr}\selectfont sbank} template bank were both $100000$ for the BNS injections, $87337$ for the NSBH injections, and $100000$ for the BBH injections.
Fig. \ref{fig:cumulative} shows the cumulative histogram of the mismatches of the combined results of the BNS, NSBH, and BBH injections for the O4 template bank and the {\fontfamily{qcr}\selectfont sbank} template bank. 
$90$ $\%$ of the injections for both banks have fitting factors of $98$ $\%$ or higher, showing that the two banks have similar effectualness.

\begin{figure*}
    \centering
    \begin{subfigure}[b]{8.6cm}
        \centering
        \includegraphics[width=8.6cm]{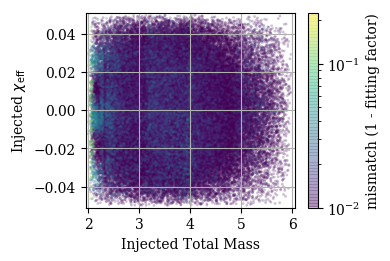}
        \caption{\protect Plot of mismatches of the BNS injections for the O4 template bank.}
        \label{fig:banksim_mani_bns_plot}
    \end{subfigure}
    \hfill
    \begin{subfigure}[b]{8.6cm}
        \centering
        \includegraphics[width=8.6cm]{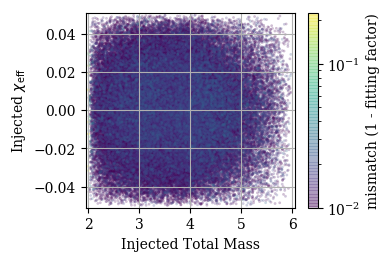}
        \caption{\protect Plot of mismatches of the BNS injections for the {\fontfamily{qcr}\selectfont sbank} template bank.}
        \label{fig:banksim_sbank_bns_plot}
    \end{subfigure}
    \caption{\protect Plots for the BNS injections}
    \medskip
    Fig. \ref{fig:banksim_mani_bns_plot} and Fig. \ref{fig:banksim_sbank_bns_plot} show the total masses of the injections on the \(x\)-axis, \(\chi_{\text{eff}}\) on the \(y\)-axis, and the mismatch (\(= 1 - \left(\text{fitting factor}\right)\)) on the color bar. To visually present the mismatch of the entire BNS parameter space, mismatches smaller than $10^{-2}$ were mapped to $10^{-2}$. $90$ $\%$ of the BNS injections have fitting factors of $98.62$ $\%$ or higher for the O4 template bank, and $98.22$ $\%$ or higher for the {\fontfamily{qcr}\selectfont sbank} template bank.
    \label{fig:banksim_bns}
\end{figure*}

\begin{figure*}
    \centering
    \begin{subfigure}[b]{8.6cm}
        \centering
        \includegraphics[width=8.6cm]{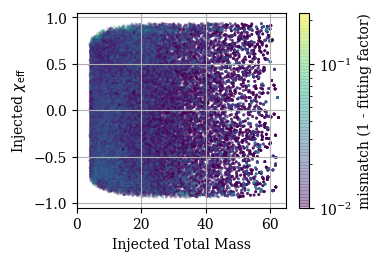}
        \caption{\protect Plot of mismatches of the NSBH injections for the O4 template bank.}
        \label{fig:banksim_mani_nsbh_plot}
    \end{subfigure}
    \hfill
    \begin{subfigure}[b]{8.6cm}
        \centering
        \includegraphics[width=8.6cm]{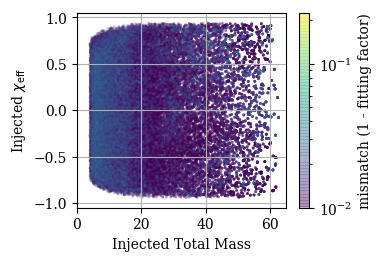}
        \caption{\protect Plot of mismatches of the NSBH injections for the {\fontfamily{qcr}\selectfont sbank} template bank.}
        \label{fig:banksim_sbank_nsbh_plot}
    \end{subfigure}
    \caption{\protect Plots for the NSBH injections}
    \medskip
    Fig. \ref{fig:banksim_mani_nsbh_plot} and Fig. \ref{fig:banksim_sbank_nsbh_plot} show the total masses of the injections on the \(x\)-axis, \(\chi_{\text{eff}}\) on the \(y\)-axis, and the mismatch (\(= 1 - \left(\text{fitting factor}\right)\)) on the color bar. To visually present the mismatch of the entire NSBH parameter space, mismatches smaller than $10^{-2}$ were mapped to $10^{-2}$. $90$ $\%$ of the NSBH injections have fitting factors of $98.17$ $\%$ or higher for the O4 template bank, and $98.15$ $\%$ or higher for the {\fontfamily{qcr}\selectfont sbank} template bank.
    \label{fig:banksim_nsbh}
\end{figure*}

\begin{figure*}
    \centering
    \begin{subfigure}[b]{8.6cm}
        \centering
        \includegraphics[width=8.6cm]{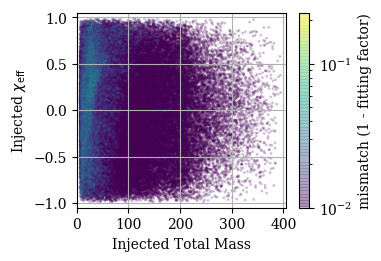}
        \caption{\protect Plot of mismatches of the BBH injections for the O4 template bank.}
        \label{fig:banksim_mani_bbh_plot}
    \end{subfigure}
    \hfill
    \begin{subfigure}[b]{8.6cm}
        \centering
        \includegraphics[width=8.6cm]{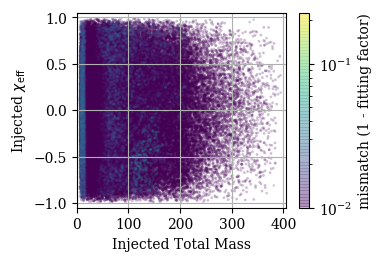}
        \caption{\protect Plot of mismatches of the BBH injections for the {\fontfamily{qcr}\selectfont sbank} template bank.}
        \label{fig:banksim_sbank_bbh_plot}
    \end{subfigure}
    \caption{\protect Plots for the BBH injections}
    \medskip
    Fig. \ref{fig:banksim_mani_bbh_plot} and Fig. \ref{fig:banksim_sbank_bbh_plot} show the total masses of the injections on the \(x\)-axis, \(\chi_{\text{eff}}\) on the \(y\)-axis, and the mismatch (\(= 1 - \left(\text{fitting factor}\right)\)) on the color bar. To visually present the mismatch of the entire BBH parameter space, mismatches smaller than $10^{-2}$ were mapped to $10^{-2}$. $90$ $\%$ of the BBH injections have fitting factors of $98.60$ $\%$ or higher for the O4 template bank, and $98.95$ $\%$ or higher for the {\fontfamily{qcr}\selectfont sbank} template bank.
    \label{fig:banksim_bbh}
\end{figure*}

Figures in Fig. \ref{fig:banksim_bns}, Fig. \ref{fig:banksim_nsbh}, and Fig. \ref{fig:banksim_bbh} show the bank simulation results for the BNS, NSBH, and BBH parameter space injections, respectively.
$90$ $\%$ of the BNS, NSBH, and BBH injections have fitting factors of $98$ $\%$ or higher for both template banks. 
Fig. \ref{fig:banksim_bns} and Fig. \ref{fig:banksim_nsbh} show that the large mismatches between the injections with the template bank are at the edge of the injected $M$ - injected $\chi_{\text{eff}}$ parameter space. 
Although bank simulation tests showed that $99$ $\%$ of the injections have fitting factors of $97$ $\%$ or higher, which is the target minimal match, the current template placing algorithm of {\fontfamily{qcr}\selectfont manifold} is not covering the edge of the template bank robustly as compared to {\fontfamily{qcr}\selectfont sbank}'s algorithm. 
Therefore, the O4 template bank generated with {\fontfamily{qcr}\selectfont manifold} exhibits higher mismatches at the boundaries of the bank. 
Improving the template placement algorithm for the boundaries of the bank is for future work. 
Fig. \ref{fig:banksim_mani_bbh_plot} shows that the mismatches between injections with the O4 template bank templates decrease as the total mass increases, reaching below $1$ $\%$ mismatches for injections above $M \geq 200 M_\odot$. 
The lower mismatches for larger mass injections are due to {\fontfamily{qcr}\selectfont manifold} requiring an upper bound as expressed in Eq. \ref{eq:manifold_coord_requirement}.  

The bank simulation results show that the O4 template bank generated with {\fontfamily{qcr}\selectfont manifold} is populated as designed and is sufficient to detect GW signals in the template bank parameter space. 
In addition to using injection generated with the same waveform approximant as the templates in the bank, authors conducted bank simulation tests using injections generated with the IMRPhenomXAS ~\cite{Pratten_2020} waveform approximant to test the template bank's performance on injections with different waveform approximants than the bank. 
The bank simulation test using the IMRPhenomXAS waveform approximant had $10^5$ injections each for the BNS, NSBH, and BBH parameter spaces, and performed as well as bank simulation tests using the IMRPhenomD waveform approximant injections. 
Therefore, the template bank generated with {\fontfamily{qcr}\selectfont manifold} is as effective as the template bank generated with {\fontfamily{qcr}\selectfont sbank}.

\subsubsection{Bank simulation for the checkerboarded template banks}
\label{subsubsection:banksim4}

In this section, we present the bank simulation results for the two checkerboarded template banks to assess the effectualness of each checkerboarded template bank when they are deployed in the online analysis. 
$100000$, $87337$, and $100000$ injections were used for BNS, NSBH, and BBH, respectively, for both template banks. 

\begin{figure*}
    \centering
    \begin{subfigure}[b]{8.6cm}
        \centering
        \includegraphics[width=8.6cm]{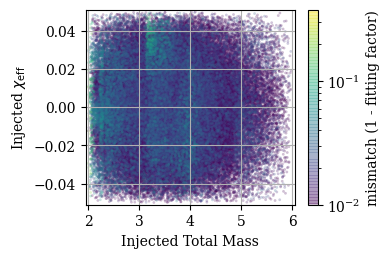}
        \caption{\protect Mismatches of the BNS injections for the left half of the O4 template bank.}
        \label{fig:banksim4_ed_bns_plot}
    \end{subfigure}
    \hfill
    \begin{subfigure}[b]{8.6cm}
        \centering
        \includegraphics[width=8.6cm]{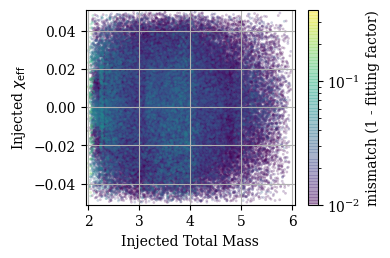}
        \caption{\protect Mismatches of the BNS injections for the right half of the O4 template bank.}
        \label{fig:banksim4_jb_bns_plot}
    \end{subfigure}
    \caption{\protect Plots for the BNS injections for the checkerboarded template banks}
    \medskip
    Fig. \ref{fig:banksim4_ed_bns_plot} and Fig. \ref{fig:banksim4_jb_bns_plot} show the total masses of the injections on the \(x\)-axis, \(\chi_{\text{eff}}\) on the \(y\)-axis, and the mismatch (\(= 1 - \left(\text{fitting factor}\right)\)) on the color bar. Lower masses have larger mismatches. To visually present the mismatch of the entire BNS parameter space, mismatches smaller than $10^{-2}$ were mapped to $10^{-2}$. $90$ $\%$ of the BNS injections have fitting factors of $97.60$ $\%$ or higher for the left half of the O4 template bank and $97.58$ $\%$ or higher for the right side of the O4 template bank.
    \label{fig:banksim4_bns}
\end{figure*}

\begin{figure*}
    \centering
    \begin{subfigure}[b]{8.6cm}
        \centering
        \includegraphics[width=8.6cm]{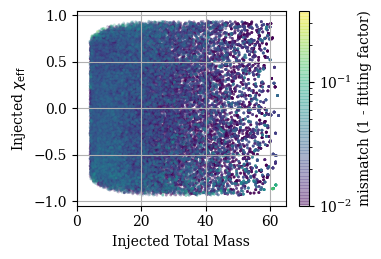}
        \caption{\protect Mismatches of the NSBH injections for the left half of the O4 template bank.}
        \label{fig:banksim4_ed_nsbh_plot}
    \end{subfigure}
    \hfill
    \begin{subfigure}[b]{8.6cm}
        \centering
        \includegraphics[width=8.6cm]{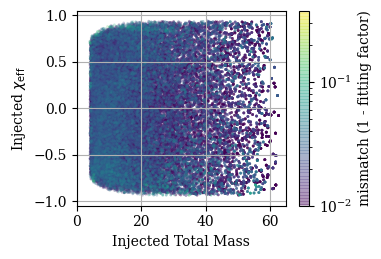}
        \caption{\protect Mismatches of the NSBH injections for the right half of the O4 template bank.}
        \label{fig:banksim4_jb_nsbh_plot}
    \end{subfigure}
    \caption{\protect Plots for the NSBH injections for the checkerboarded template banks}
    \medskip
    Fig. \ref{fig:banksim4_ed_nsbh_plot} and Fig. \ref{fig:banksim4_jb_nsbh_plot} show the total masses of the injections on the \(x\)-axis, \(\chi_{\text{eff}}\) on the \(y\)-axis, and the mismatch (\(= 1 - \left(\text{fitting factor}\right)\)) on the color bar. Lower masses have higher mismatches. To visually present the mismatch of the entire NSBH parameter space, mismatches smaller than $10^{-2}$ were mapped to $10^{-2}$. $90$ $\%$ of the NSBH injections have fitting factors of $97.07$ $\%$ or higher for the left half of the O4 template bank and $97.10$ $\%$ or higher for the right side of the O4 template bank.
    \label{fig:banksim4_nsbh}
\end{figure*}

\begin{figure*}
    \centering
    \begin{subfigure}[b]{8.6cm}
        \centering
        \includegraphics[width=8.6cm]{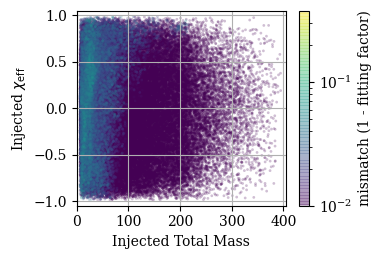}
        \caption{\protect Mismatches of the BBH injections for the left half of the O4 template bank.}
        \label{fig:banksim4_ed_bbh_plot}
    \end{subfigure}
    \hfill
    \begin{subfigure}[b]{8.6cm}
        \centering
        \includegraphics[width=8.6cm]{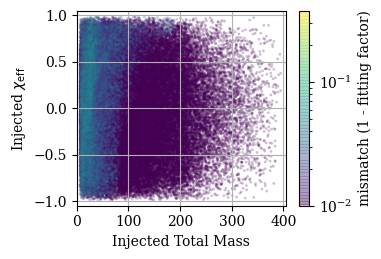}
        \caption{\protect Mismatches of the BBH injections for the right half of the O4 template bank.}
        \label{fig:banksim4_jb_bbh_plot}
    \end{subfigure}
    \caption{\protect Plots for the BBH injections for the checkerboarded template banks}
    \medskip
    Fig. \ref{fig:banksim4_ed_bbh_plot} and Fig. \ref{fig:banksim4_jb_bbh_plot} show the total masses of the injections on the \(x\)-axis, \(\chi_{\text{eff}}\) on the \(y\)-axis, and the mismatch (\(= 1 - \left(\text{fitting factor}\right)\)) on the color bar. Lower masses have larger mismatches. To visually present the mismatch of the entire BBH parameter space, mismatches smaller than $10^{-2}$ were mapped to $10^{-2}$. $90$ $\%$ of the BBH injections have fitting factors of $97.82$ $\%$ or higher for the left half of the O4 template bank and $97.83$ $\%$ or higher for the right side of the O4 template bank.
    \label{fig:banksim4_bbh}
\end{figure*}

Fig. \ref{fig:banksim4_bns}, Fig. \ref{fig:banksim4_nsbh} and Fig. \ref{fig:banksim4_bbh} show the bank simulation results for the BNS, NSBH, and BBH parameter space injections for the checkerboarded template banks and show that the bulk of the injections has fitting factors above $97$ $\%$, and fitting factors are low at the edges of the checkerboarded template banks in the $M$ - $\chi_{\text{eff}}$ parameter space, which is a behavior that the O4 template bank exhibits as well. 
The BNS, NSBH, and BBH injections have fitting factors of $\sim 98$ $\%$, $\sim 97$ $\%$, and $\sim 98$ $\%$, respectively, for $90$ $\%$ of the injections in each source category, which are $\sim 1$ $\%$ less than that of the pre-checkerboarded O4 template bank. 
The low fitting factors in Fig. \ref{fig:banksim4_bns} for the BNS injections can be attributed to injections that were at the edge of the template bank in the $m_1$ - $m_2$ space. 
While the checkerboarded template banks have lower fitting factors than the pre-checkerboarded O4 template bank, each checkerboarded template bank still maintains a minimum of a $97$ $\%$ fitting factor for $90$ $\%$ of the injections and thus are effective to detect GW signals in the online analyses in cases that one of the checkerboarded template banks cannot process data. 

\section{Conclusion}
\label{section:conclusion}
We have presented the design and tests of the template bank used by {\fontfamily{qcr}\selectfont GstLAL} to analyze data from the fourth observing run of Advanced LIGO, Virgo, and KAGRA. 
In contrast to the previous template banks generated for the {\fontfamily{qcr}\selectfont GstLAL} searches in O1, O2, and O3, the template bank we present is generated using a computationally efficient binary tree approach of template placements,  {\fontfamily{qcr}\selectfont manifold}, instead of a stochastic template placement, {\fontfamily{qcr}\selectfont LALApps} {\fontfamily{qcr}\selectfont sbank}. 
For the O4 template bank, we applied a new SVD sorting scheme that implements PN phase terms and improves the SVD efficiency by nearly $5$ times compared to the previously used SVD sorting schemes. 
As the SVD sorting parameters were originally intended for lower mass systems, there is potential for improvement.
The O4 template bank generated with {\fontfamily{qcr}\selectfont manifold} spans the mass parameter space of $1$ - $200$ $M_\odot$ in component mass, total mass of $2$ - $400$ $M_\odot$, the dimensionless spins are $\pm 0.05$ for component masses $1$ - $3$ $M_\odot$ and $\pm 0.99$ for component masses $3$ - $200$ $M_\odot$, and the mass ratios ranges between $1$ - $20$.
The LIGO O4 PSD for simulation purposes was used and the low-frequency cutoff was chosen to be $10 \mathrm{Hz}$ with a waveform maximum duration cut of $128$ seconds, and the lowest frequency for each template depends on its duration.  
The bank simulation tests have shown that the O4 template bank is as effective as the template bank generated with {\fontfamily{qcr}\selectfont LALApps} {\fontfamily{qcr}\selectfont sbank} for the same parameter space, and that both checkerboarded template banks deployed in the online analysis have fitting factors above $97$ $\%$ for $90$ $\%$ of the injections across the entire O4 template bank parameter space.
Thus, the O4 template bank presented in this paper is sufficient to detect GW signals from BNS, NSBH, and BBH events up to a total mass of $400$ $M_\odot$, and checkerboarded template banks enable the {\fontfamily{qcr}\selectfont GstLAL} pipeline to detect GW signals even in situations with only one checkerboarded template bank analyzing GW data at a slightly lower fitting factor than when using the entire template bank. 
The O4 template bank has lower fitting factors at the edges of the template bank parameter space as compared to the rest of the template bank parameter space, thus the algorithmic changes to {\fontfamily{qcr}\selectfont manifold} will be future work. 
In O4, the same template bank will be used for both the online and the offline analysis for consistency of the analyses. 


\begin{acknowledgments}
We thank the LIGO-Virgo-KAGRA Scientific Collaboration for access to data.
This material is based upon work supported by NSF's LIGO Laboratory which is a major facility fully funded by the National Science Foundation.
LIGO was constructed by the California Institute of Technology and Massachusetts Institute of Technology with funding from the National Science Foundation (NSF) and operates under cooperative agreements PHYS-\(0757058\) and PHY-\(0823459\).
We thank B Sathyaprakash and Elisa Nitoglia for their helpful comments and suggestions. 
We also thank Graham Woan for helping with the review of the O4 template bank. 
The authors are grateful for the computational resources provided by the LIGO Laboratory and supported by National Science Foundation Grants PHY-$0757058$ and PHY-$0823459$, and the Pennsylvania State University's Institute for Computational and Data Sciences (ICDS) and supported by NSF PHY-\(2011865\) and NSF OAC-\(2103662\), and the University of Wisconsin Milwaukee Nemo and support by NSF PHY-\(1626190\), NSF PHY-\(1700765\), and NSF PHY-\(2207728\).
H.F. was supported by the JSPS Postdoctoral Fellowships for Research in Japan. 

\end{acknowledgments}

\appendix
\section{Bank simulations with injections outside the template bank parameter space}
\label{section:appendix}

The O4 template bank's parameters are chosen to be what is illustrated in section ~\ref{section:design} to limit computational costs while the authors acknowledge that there could be signals that lie outside these limits.
Bank simulation tests described below are performed to understand the O4 template bank's performance on such signals.  
Authors performed two bank simulation tests on the O4 template bank and the {\fontfamily{qcr}\selectfont sbank} template bank using injections that lie outside the template bank parameter space that were generated using the same waveform approximant and same PSD as the template banks:
\begin{enumerate}[label=\Alph*.]
    \item Assess the behavior of the O4 template bank and compare with that of the {\fontfamily{qcr}\selectfont sbank} template bank using injections that span NS spins from $-0.99$ to $0.99$ for the BNS and NSBH parameter space injections. (Section \ref{subsection:banksim2})
    \item Assess the behavior of the O4 template bank and compare with that of the {\fontfamily{qcr}\selectfont sbank} template bank using injections that span mass ratios up to $q = 50$ for the NSBH and BBH parameter space injections. (Section \ref{subsection:banksim3})
\end{enumerate}
The bank simulation tests in this section show that the template banks generated with {\fontfamily{qcr}\selectfont manifold} and {\fontfamily{qcr}\selectfont sbank} performed similarly.

\subsection{Bank simulation with large NS spin injections}
\label{subsection:banksim2}

NSs in binary systems have been observed to have spins to be \(0.04\) at largest ~\cite{Kramer_2009, Burgay_2003, Stovall_2018}, which motivated the O4 template bank to span \(-0.05\) to \(0.05\) for NS spins. 
Meanwhile, NSs with spins as large as \(0.4\) have been observed ~\cite{Hessels_2006}, and sets of NS EOS studied in ~\cite{Lo_2011} go up to \(0.7\) for NS spins. 
Therefore, the purpose of the bank simulations described in this section is to test the template banks against systems that have NSs with spins larger (smaller) than \(0.05\) (\(-0.05\)). 

\begin{figure*}
    \centering
    \begin{subfigure}[b]{8.6cm}
        \centering
        \includegraphics[width=8.6cm]{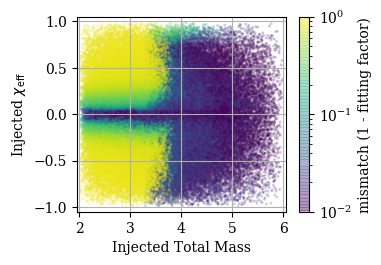}
        \caption{\protect Mismatches of the BNS injections with large NS spins for the O4 template bank.}
        \label{fig:banksim2_mani_bns_injM_injchi}
    \end{subfigure}
    \hfill
    \begin{subfigure}[b]{8.6cm}
        \centering
        \includegraphics[width=8.6cm]{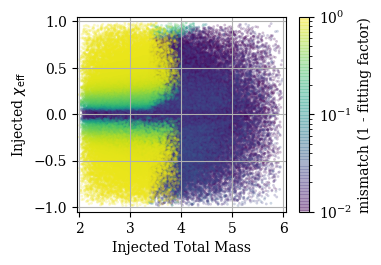}
        \caption{\protect Mismatches of the BNS injections with large NS spins for the  {\fontfamily{qcr}\selectfont sbank} template bank.}
        \label{fig:banksim2_sbank_bns_injM_injchi}
    \end{subfigure}
    \caption{\protect Plots for the BNS region injections with large NS spins}
    \medskip
    Fig. \ref{fig:banksim2_mani_bns_injM_injchi} and Fig. \ref{fig:banksim2_sbank_bns_injM_injchi} show the total masses of the injections on the \(x\)-axis, \(\chi_{\text{eff}}\) on the \(y\)-axis, and the mismatch (\(= 1 - \left(\text{fitting factor}\right)\)) on the color bar. To visually present the mismatch of the entire BNS parameter space, mismatches smaller than $10^{-2}$ were mapped to $10^{-2}$. $90$ $\%$ of the BNS injections with large NS spins have fitting factors of $16.28$ $\%$ or higher for the O4 template bank, and $15.55$ $\%$ or higher for the {\fontfamily{qcr}\selectfont sbank} template bank.
    \label{fig:banksim2_bns}
\end{figure*}

\begin{figure*}
    \centering
    \begin{subfigure}[b]{8.6cm}
        \centering
        \includegraphics[width=8.6cm]{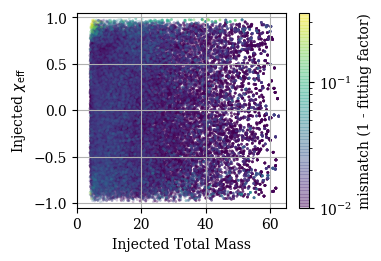}
        \caption{\protect Mismatches of the NSBH injections with large NS spins for the O4 template bank.}
        \label{fig:banksim2_mani_nsbh_injM_injchi}
    \end{subfigure}
    \hfill
    \begin{subfigure}[b]{8.6cm}
        \centering
        \includegraphics[width=8.6cm]{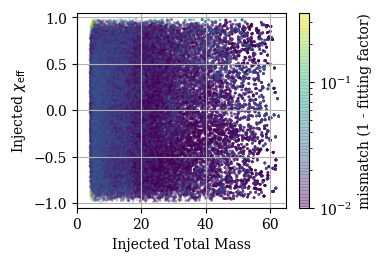}
        \caption{\protect Mismatches of the NSBH injections with large NS spins for the  {\fontfamily{qcr}\selectfont sbank} template bank.}
        \label{fig:banksim2_sbank_nsbh_injM_injchi}
    \end{subfigure}
    \caption{\protect Plots for the NSBH injections with large NS spins}
    \medskip
    Fig. \ref{fig:banksim2_mani_nsbh_injM_injchi} and Fig. \ref{fig:banksim2_sbank_nsbh_injM_injchi} show the total masses of the injections on the \(x\)-axis, \(\chi_{\text{eff}}\) on the \(y\)-axis, and the mismatch (\(= 1 - \left(\text{fitting factor}\right)\)) on the color bar. To visually present the mismatch of the entire NSBH parameter space, mismatches smaller than $10^{-2}$ were mapped to $10^{-2}$.
    The mismatches are larger as the injection $\chi_{\text{eff}}$ values approach the upper (lower) limits. $90$ $\%$ of the NSBH injections with large NS spins have fitting factors of $98.13$ $\%$ or higher for the O4 template bank, and $98.12$ $\%$ or higher for the {\fontfamily{qcr}\selectfont sbank} template bank.
    \label{fig:banksim2_nsbh}
\end{figure*}

The numbers of injections used for the O4 template bank and the {\fontfamily{qcr}\selectfont sbank} template bank were $100000$ for BNS injections and $87337$ for NSBH injections.
Figures in Fig. \ref{fig:banksim2_bns} and Fig. \ref{fig:banksim2_nsbh} show the bank simulation results for the BNS and NSBH parameter space injections, respectively.
These injection sets have injections with NS spins spanning \(-0.99\) to \(0.99\) for both the O4 template bank generated with {\fontfamily{qcr}\selectfont manifold} and the {\fontfamily{qcr}\selectfont sbank} template bank.
Both template banks have comparable effectiveness against the BNS and NSBH injections with large NS spins, and the mismatches are larger as the injection \(|\chi_{\text{eff}}|\) values increase. 

\subsection{Bank simulation with large mass ratio injections}
\label{subsection:banksim3}

This choice of setting the maximum mass ratio to \(20\) was motivated by the range of calibration of the waveform models ~\cite{Ossokine_2020} and by detections of GW events in O3 ~\cite{theligoscientificcollaboration2021gwtc3}. 
Meanwhile, events like \(\text{GW}191219\_163120\) have mass ratios larger than \(q = 20\) ~\cite{theligoscientificcollaboration2021gwtc3}. 
The purpose of the bank simulation results presented in this section is to test the effectiveness of the template banks against injections with mass ratios up to \(q = 50\). 

\begin{figure*}
    \centering
    \begin{subfigure}[b]{8.6cm}
        \centering
        \includegraphics[width=8.6cm]{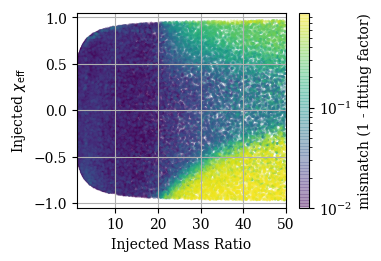}
        \caption{\protect Mismatches of the NSBH injections with mass ratios up to \(50\) for the O4 template bank.}
        \label{fig:banksim3_mani_nsbh_q_chi}
    \end{subfigure}
    \hfill
    \begin{subfigure}[b]{8.6cm}
        \centering
        \includegraphics[width=8.6cm]{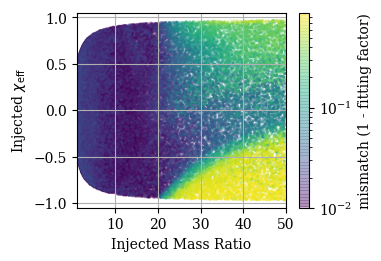}
        \caption{\protect Mismatches of the NSBH injections with mass ratios up to \(50\) for {\fontfamily{qcr}\selectfont sbank} template bank.}
        \label{fig:banksim3_sbank_nsbh_q_chi}
    \end{subfigure}
    \caption{\protect Plots for the NSBH injections with mass ratios up to \(q = 50\)}
    \medskip
    Fig. \ref{fig:banksim3_mani_nsbh_q_chi} and Fig. \ref{fig:banksim3_sbank_nsbh_q_chi} show the mass ratio of the injections on the \(x\)-axis, \(\chi_{\text{eff}}\) on the \(y\)-axis, and the mismatch (\(= 1 - \left(\text{fitting factor}\right)\)) on the color bar. The figures show that large mismatches occur for large positive (negative values) of $\chi_{\text{eff}}$ and for mass ratios above $20$. To visually present the mismatch of the entire NSBH parameter space, mismatches smaller than $10^{-2}$ were mapped to $10^{-2}$. $90$ $\%$ of the NSBH injections with mass ratio up to $50$ have fitting factors of $80.53$ $\%$ or higher for the O4 template bank, and $78.44$ $\%$ or higher for the {\fontfamily{qcr}\selectfont sbank} template bank.
    \label{fig:banksim3_nsbh}
\end{figure*}

\begin{figure*}
    \centering
    \begin{subfigure}[b]{8.6cm}
        \centering
        \includegraphics[width=8.6cm]{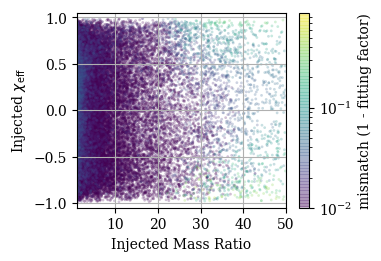}
        \caption{\protect Mismatches of the BBH injections with mass ratios up to \(50\) for the O4 template bank.}
        \label{fig:banksim3_mani_bbh_q_chi}
    \end{subfigure}
    \hfill
    \begin{subfigure}[b]{8.6cm}
        \centering
        \includegraphics[width=8.6cm]{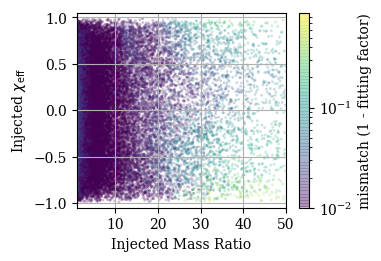}
        \caption{\protect Mismatches of the BBH injections with mass ratios up to \(50\) for {\fontfamily{qcr}\selectfont sbank} template bank.}
        \label{fig:banksim3_sbank_bbh_q_chi}
    \end{subfigure}
    \caption{\protect Plots for the BBH injections with mass ratio up to \(q = 50\)}
    \medskip
    Fig. \ref{fig:banksim3_mani_bbh_q_chi} and Fig. \ref{fig:banksim3_sbank_bbh_q_chi} show the mass ratio of the injections on the \(x\)-axis, \(\chi_{\text{eff}}\) on the \(y\)-axis, and the mismatch (\(= 1 - \left(\text{fitting factor}\right)\)) on the color bar. The figures show that large mismatches occur for large mass ratio injections. To visually present the mismatch of the entire BBH parameter space, mismatches smaller than $10^{-2}$ were mapped to $10^{-2}$. $90$ $\%$ of the BBH injections with mass ratio up to $50$ have fitting factors of $98.34$ $\%$ or higher for the O4 template bank, and $98.39$ $\%$ or higher for the {\fontfamily{qcr}\selectfont sbank} template bank.
    \label{fig:banksim3_bbh}
\end{figure*}

The numbers of injections used for the O4 template bank and the {\fontfamily{qcr}\selectfont sbank} template bank were both $98933$ for NSBH injections and $100000$ for BBH injections.
Figures in Fig. \ref{fig:banksim3_nsbh} and Fig. \ref{fig:banksim3_bbh} show the bank simulation results for the NSBH and BBH parameter space injections with $q$ up to $50$, respectively.
The O4 template bank and {\fontfamily{qcr}\selectfont sbank} template bank have a fitting factor of above $90$ $\%$ for $\sim 80$ $\%$ of the NSBH injections, while the O4 template bank and {\fontfamily{qcr}\selectfont sbank} template bank have a fitting factor of above $90$ $\%$ for $\sim 98$ $\%$ of the BBH injections. 
Fig. \ref{fig:banksim3_nsbh} shows that the O4 template bank and the {\fontfamily{qcr}\selectfont sbank} template bank have similar responses to the NSBH injections with mass ratios up to \(q = 50\).
Fig. \ref{fig:banksim3_bbh} shows that the O4 template bank has lower mismatches for injections with $M \geq 200 M_\odot$, which is a result of applying Eq. \ref{eq:manifold_coord_requirement}. 

\bibliography{reference}

\providecommand{\noopsort}[1]{}\providecommand{\singleletter}[1]{#1}%
\begin{thebibliography}{96}%
\makeatletter
\providecommand \@ifxundefined [1]{%
 \@ifx{#1\undefined}
}%
\providecommand \@ifnum [1]{%
 \ifnum #1\expandafter \@firstoftwo
 \else \expandafter \@secondoftwo
 \fi
}%
\providecommand \@ifx [1]{%
 \ifx #1\expandafter \@firstoftwo
 \else \expandafter \@secondoftwo
 \fi
}%
\providecommand \natexlab [1]{#1}%
\providecommand \enquote  [1]{``#1''}%
\providecommand \bibnamefont  [1]{#1}%
\providecommand \bibfnamefont [1]{#1}%
\providecommand \citenamefont [1]{#1}%
\providecommand \href@noop [0]{\@secondoftwo}%
\providecommand \href [0]{\begingroup \@sanitize@url \@href}%
\providecommand \@href[1]{\@@startlink{#1}\@@href}%
\providecommand \@@href[1]{\endgroup#1\@@endlink}%
\providecommand \@sanitize@url [0]{\catcode `\\12\catcode `\$12\catcode
  `\&12\catcode `\#12\catcode `\^12\catcode `\_12\catcode `\%12\relax}%
\providecommand \@@startlink[1]{}%
\providecommand \@@endlink[0]{}%
\providecommand \url  [0]{\begingroup\@sanitize@url \@url }%
\providecommand \@url [1]{\endgroup\@href {#1}{\urlprefix }}%
\providecommand \urlprefix  [0]{URL }%
\providecommand \Eprint [0]{\href }%
\providecommand \doibase [0]{http://dx.doi.org/}%
\providecommand \selectlanguage [0]{\@gobble}%
\providecommand \bibinfo  [0]{\@secondoftwo}%
\providecommand \bibfield  [0]{\@secondoftwo}%
\providecommand \translation [1]{[#1]}%
\providecommand \BibitemOpen [0]{}%
\providecommand \bibitemStop [0]{}%
\providecommand \bibitemNoStop [0]{.\EOS\space}%
\providecommand \EOS [0]{\spacefactor3000\relax}%
\providecommand \BibitemShut  [1]{\csname bibitem#1\endcsname}%
\let\auto@bib@innerbib\@empty
\bibitem [{\citenamefont {Harry}\ \emph {et~al.}(2010)\citenamefont {Harry}
  \emph {et~al.}}]{Harry2010}%
  \BibitemOpen
  \bibfield  {author} {\bibinfo {author} {\bibfnamefont {G.~M.}\ \bibnamefont
  {Harry}} \emph {et~al.} (\bibinfo {collaboration} {LIGO Scientific}),\ }\href
  {\doibase 10.1088/0264-9381/27/8/084006} {\bibfield  {journal} {\bibinfo
  {journal} {Classical and Quantum Gravity}\ }\textbf {\bibinfo {volume}
  {27}},\ \bibinfo {pages} {084006} (\bibinfo {year} {2010})}\BibitemShut
  {NoStop}%
\bibitem [{\citenamefont {Abbott}\ \emph
  {et~al.}(2016{\natexlab{a}})\citenamefont {Abbott} \emph
  {et~al.}}]{TheLIGOScientific:2016pea}%
  \BibitemOpen
  \bibfield  {author} {\bibinfo {author} {\bibfnamefont {B.~P.}\ \bibnamefont
  {Abbott}} \emph {et~al.} (\bibinfo {collaboration} {LIGO Scientific,
  Virgo}),\ }\href {\doibase 10.1103/PhysRevX.6.041015,
  10.1103/PhysRevX.8.039903} {\bibfield  {journal} {\bibinfo  {journal} {Phys.
  Rev. X}\ }\textbf {\bibinfo {volume} {6}},\ \bibinfo {pages} {041015}
  (\bibinfo {year} {2016}{\natexlab{a}})},\ \Eprint
  {http://arxiv.org/abs/1606.04856} {arXiv:1606.04856 [gr-qc]} \BibitemShut
  {NoStop}%
\bibitem [{\citenamefont {Aasi}\ \emph {et~al.}(2015)\citenamefont {Aasi} \emph
  {et~al.}}]{TheLIGOScientific:2014jea}%
  \BibitemOpen
  \bibfield  {author} {\bibinfo {author} {\bibfnamefont {J.}~\bibnamefont
  {Aasi}} \emph {et~al.} (\bibinfo {collaboration} {LIGO Scientific}),\ }\href
  {\doibase 10.1088/0264-9381/32/7/074001} {\bibfield  {journal} {\bibinfo
  {journal} {Class. Quant. Grav.}\ }\textbf {\bibinfo {volume} {32}},\ \bibinfo
  {pages} {074001} (\bibinfo {year} {2015})},\ \Eprint
  {http://arxiv.org/abs/1411.4547} {arXiv:1411.4547 [gr-qc]} \BibitemShut
  {NoStop}%
\bibitem [{\citenamefont {Abbott}\ \emph
  {et~al.}(2019{\natexlab{a}})\citenamefont {Abbott} \emph
  {et~al.}}]{LIGOScientific:2018mvr}%
  \BibitemOpen
  \bibfield  {author} {\bibinfo {author} {\bibfnamefont {B.~P.}\ \bibnamefont
  {Abbott}} \emph {et~al.} (\bibinfo {collaboration} {LIGO Scientific,
  Virgo}),\ }\href {\doibase 10.1103/PhysRevX.9.031040} {\bibfield  {journal}
  {\bibinfo  {journal} {Phys. Rev. X}\ }\textbf {\bibinfo {volume} {9}},\
  \bibinfo {pages} {031040} (\bibinfo {year} {2019}{\natexlab{a}})},\ \Eprint
  {http://arxiv.org/abs/1811.12907} {arXiv:1811.12907 [astro-ph.HE]}
  \BibitemShut {NoStop}%
\bibitem [{\citenamefont {Abbott}\ \emph
  {et~al.}(2016{\natexlab{b}})\citenamefont {Abbott} \emph
  {et~al.}}]{Abbott:2016blz}%
  \BibitemOpen
  \bibfield  {author} {\bibinfo {author} {\bibfnamefont {B.~P.}\ \bibnamefont
  {Abbott}} \emph {et~al.} (\bibinfo {collaboration} {LIGO Scientific,
  Virgo}),\ }\href {\doibase 10.1103/PhysRevLett.116.061102} {\bibfield
  {journal} {\bibinfo  {journal} {Phys. Rev. Lett.}\ }\textbf {\bibinfo
  {volume} {116}},\ \bibinfo {pages} {061102} (\bibinfo {year}
  {2016}{\natexlab{b}})},\ \Eprint {http://arxiv.org/abs/1602.03837}
  {arXiv:1602.03837 [gr-qc]} \BibitemShut {NoStop}%
\bibitem [{\citenamefont {{The LIGO Scientific Collaboration}}\ \emph
  {et~al.}(2022)\citenamefont {{The LIGO Scientific Collaboration}},
  \citenamefont {{the Virgo Collaboration}}, \citenamefont {Abbott} \emph
  {et~al.}}]{gwtc2_1}%
  \BibitemOpen
  \bibfield  {author} {\bibinfo {author} {\bibnamefont {{The LIGO Scientific
  Collaboration}}}, \bibinfo {author} {\bibnamefont {{the Virgo
  Collaboration}}}, \bibinfo {author} {\bibfnamefont {R.}~\bibnamefont
  {Abbott}},  \emph {et~al.},\ }\href@noop {} {\enquote {\bibinfo {title}
  {{GWTC-2.1: Deep Extended Catalog of Compact Binary Coalescences Observed by
  LIGO and Virgo During the First Half of the Third Observing Run}},}\ }
  (\bibinfo {year} {2022}),\ \Eprint {http://arxiv.org/abs/2108.01045}
  {arXiv:2108.01045 [gr-qc]} \BibitemShut {NoStop}%
\bibitem [{\citenamefont {Abbott}\ \emph
  {et~al.}(2016{\natexlab{c}})\citenamefont {Abbott} \emph
  {et~al.}}]{Abbott:2016nmj}%
  \BibitemOpen
  \bibfield  {author} {\bibinfo {author} {\bibfnamefont {B.~P.}\ \bibnamefont
  {Abbott}} \emph {et~al.} (\bibinfo {collaboration} {LIGO Scientific,
  Virgo}),\ }\href {\doibase 10.1103/PhysRevLett.116.241103} {\bibfield
  {journal} {\bibinfo  {journal} {Phys. Rev. Lett.}\ }\textbf {\bibinfo
  {volume} {116}},\ \bibinfo {pages} {241103} (\bibinfo {year}
  {2016}{\natexlab{c}})},\ \Eprint {http://arxiv.org/abs/1606.04855}
  {arXiv:1606.04855 [gr-qc]} \BibitemShut {NoStop}%
\bibitem [{\citenamefont {Abbott}\ \emph
  {et~al.}(2019{\natexlab{b}})\citenamefont {Abbott} \emph
  {et~al.}}]{LIGOScientific:2019fpa}%
  \BibitemOpen
  \bibfield  {author} {\bibinfo {author} {\bibfnamefont {B.~P.}\ \bibnamefont
  {Abbott}} \emph {et~al.} (\bibinfo {collaboration} {LIGO Scientific,
  Virgo}),\ }\href {\doibase 10.1103/PhysRevD.100.104036} {\bibfield  {journal}
  {\bibinfo  {journal} {Phys. Rev. D}\ }\textbf {\bibinfo {volume} {100}},\
  \bibinfo {pages} {104036} (\bibinfo {year} {2019}{\natexlab{b}})},\ \Eprint
  {http://arxiv.org/abs/1903.04467} {arXiv:1903.04467 [gr-qc]} \BibitemShut
  {NoStop}%
\bibitem [{\citenamefont {Acernese}\ \emph {et~al.}(2015)\citenamefont
  {Acernese} \emph {et~al.}}]{TheVirgo:2014hva}%
  \BibitemOpen
  \bibfield  {author} {\bibinfo {author} {\bibfnamefont {F.}~\bibnamefont
  {Acernese}} \emph {et~al.} (\bibinfo {collaboration} {VIRGO}),\ }\href
  {\doibase 10.1088/0264-9381/32/2/024001} {\bibfield  {journal} {\bibinfo
  {journal} {Class. Quant. Grav.}\ }\textbf {\bibinfo {volume} {32}},\ \bibinfo
  {pages} {024001} (\bibinfo {year} {2015})},\ \Eprint
  {http://arxiv.org/abs/1408.3978} {arXiv:1408.3978 [gr-qc]} \BibitemShut
  {NoStop}%
\bibitem [{\citenamefont {Abbott}\ \emph
  {et~al.}(2019{\natexlab{c}})\citenamefont {Abbott} \emph
  {et~al.}}]{Abbott:2018hgk}%
  \BibitemOpen
  \bibfield  {author} {\bibinfo {author} {\bibfnamefont {B.~P.}\ \bibnamefont
  {Abbott}} \emph {et~al.} (\bibinfo {collaboration} {LIGO Scientific,
  Virgo}),\ }\href {\doibase 10.3847/1538-4357/ab0f3d} {\bibfield  {journal}
  {\bibinfo  {journal} {Astrophys. J.}\ }\textbf {\bibinfo {volume} {875}},\
  \bibinfo {pages} {160} (\bibinfo {year} {2019}{\natexlab{c}})},\ \Eprint
  {http://arxiv.org/abs/1810.02581} {arXiv:1810.02581 [gr-qc]} \BibitemShut
  {NoStop}%
\bibitem [{\citenamefont {Abbott}\ \emph
  {et~al.}(2019{\natexlab{d}})\citenamefont {Abbott} \emph
  {et~al.}}]{Abbott:2018wiz}%
  \BibitemOpen
  \bibfield  {author} {\bibinfo {author} {\bibfnamefont {B.~P.}\ \bibnamefont
  {Abbott}} \emph {et~al.} (\bibinfo {collaboration} {LIGO Scientific,
  Virgo}),\ }\href {\doibase 10.1103/PhysRevX.9.011001} {\bibfield  {journal}
  {\bibinfo  {journal} {Phys. Rev. X}\ }\textbf {\bibinfo {volume} {9}},\
  \bibinfo {pages} {011001} (\bibinfo {year} {2019}{\natexlab{d}})},\ \Eprint
  {http://arxiv.org/abs/1805.11579} {arXiv:1805.11579 [gr-qc]} \BibitemShut
  {NoStop}%
\bibitem [{\citenamefont {Abbott}\ \emph
  {et~al.}(2017{\natexlab{a}})\citenamefont {Abbott} \emph
  {et~al.}}]{TheLIGOScientific:2017qsa}%
  \BibitemOpen
  \bibfield  {author} {\bibinfo {author} {\bibfnamefont {B.~P.}\ \bibnamefont
  {Abbott}} \emph {et~al.} (\bibinfo {collaboration} {LIGO Scientific,
  Virgo}),\ }\href {\doibase 10.1103/PhysRevLett.119.161101} {\bibfield
  {journal} {\bibinfo  {journal} {Phys. Rev. Lett.}\ }\textbf {\bibinfo
  {volume} {119}},\ \bibinfo {pages} {161101} (\bibinfo {year}
  {2017}{\natexlab{a}})},\ \Eprint {http://arxiv.org/abs/1710.05832}
  {arXiv:1710.05832 [gr-qc]} \BibitemShut {NoStop}%
\bibitem [{\citenamefont {Abbott}\ \emph
  {et~al.}(2017{\natexlab{b}})\citenamefont {Abbott} \emph
  {et~al.}}]{Abbott:2017dke}%
  \BibitemOpen
  \bibfield  {author} {\bibinfo {author} {\bibfnamefont {B.~P.}\ \bibnamefont
  {Abbott}} \emph {et~al.} (\bibinfo {collaboration} {LIGO Scientific,
  Virgo}),\ }\href {\doibase 10.3847/2041-8213/aa9a35} {\bibfield  {journal}
  {\bibinfo  {journal} {Astrophys. J.}\ }\textbf {\bibinfo {volume} {851}},\
  \bibinfo {pages} {L16} (\bibinfo {year} {2017}{\natexlab{b}})},\ \Eprint
  {http://arxiv.org/abs/1710.09320} {arXiv:1710.09320 [astro-ph.HE]}
  \BibitemShut {NoStop}%
\bibitem [{\citenamefont {Abbott}\ \emph
  {et~al.}(2017{\natexlab{c}})\citenamefont {Abbott} \emph
  {et~al.}}]{Monitor:2017mdv}%
  \BibitemOpen
  \bibfield  {author} {\bibinfo {author} {\bibfnamefont {B.~P.}\ \bibnamefont
  {Abbott}} \emph {et~al.} (\bibinfo {collaboration} {LIGO Scientific, Virgo,
  Fermi-GBM, INTEGRAL}),\ }\href {\doibase 10.3847/2041-8213/aa920c} {\bibfield
   {journal} {\bibinfo  {journal} {Astrophys. J.}\ }\textbf {\bibinfo {volume}
  {848}},\ \bibinfo {pages} {L13} (\bibinfo {year} {2017}{\natexlab{c}})},\
  \Eprint {http://arxiv.org/abs/1710.05834} {arXiv:1710.05834 [astro-ph.HE]}
  \BibitemShut {NoStop}%
\bibitem [{\citenamefont {Abbott}\ \emph
  {et~al.}(2017{\natexlab{d}})\citenamefont {Abbott} \emph
  {et~al.}}]{Abbott_2017}%
  \BibitemOpen
  \bibfield  {author} {\bibinfo {author} {\bibfnamefont {B.~P.}\ \bibnamefont
  {Abbott}} \emph {et~al.},\ }\href {\doibase 10.3847/2041-8213/aa91c9}
  {\bibfield  {journal} {\bibinfo  {journal} {The Astrophysical Journal
  Letters}\ }\textbf {\bibinfo {volume} {848}},\ \bibinfo {pages} {L12}
  (\bibinfo {year} {2017}{\natexlab{d}})}\BibitemShut {NoStop}%
\bibitem [{\citenamefont {{The LIGO Scientific Collaboration}}\ \emph
  {et~al.}(2021)\citenamefont {{The LIGO Scientific Collaboration}},
  \citenamefont {{the Virgo Collaboration}}, \citenamefont {{the KAGRA
  Collaboration}}, \citenamefont {Abbott} \emph
  {et~al.}}]{theligoscientificcollaboration2021gwtc3}%
  \BibitemOpen
  \bibfield  {author} {\bibinfo {author} {\bibnamefont {{The LIGO Scientific
  Collaboration}}}, \bibinfo {author} {\bibnamefont {{the Virgo
  Collaboration}}}, \bibinfo {author} {\bibnamefont {{the KAGRA
  Collaboration}}}, \bibinfo {author} {\bibfnamefont {R.}~\bibnamefont
  {Abbott}},  \emph {et~al.},\ }\href@noop {} {\enquote {\bibinfo {title}
  {{{GWTC-3}: Compact Binary Coalescences Observed by LIGO and Virgo During the
  Second Part of the Third Observing Run}},}\ } (\bibinfo {year} {2021}),\
  \Eprint {http://arxiv.org/abs/2111.03606} {arXiv:2111.03606 [gr-qc]}
  \BibitemShut {NoStop}%
\bibitem [{\citenamefont {Abbott}\ \emph
  {et~al.}(2021{\natexlab{a}})\citenamefont {Abbott} \emph
  {et~al.}}]{Abbott_2021}%
  \BibitemOpen
  \bibfield  {author} {\bibinfo {author} {\bibfnamefont {R.}~\bibnamefont
  {Abbott}} \emph {et~al.} (\bibinfo {collaboration} {LIGO Scientific, Virgo,
  KAGRA}),\ }\href {\doibase 10.3847/2041-8213/ac082e} {\bibfield  {journal}
  {\bibinfo  {journal} {The Astrophysical Journal Letters}\ }\textbf {\bibinfo
  {volume} {915}},\ \bibinfo {pages} {L5} (\bibinfo {year}
  {2021}{\natexlab{a}})}\BibitemShut {NoStop}%
\bibitem [{\citenamefont {Collaboration}()}]{gracedb_public_o4}%
  \BibitemOpen
  \bibfield  {author} {\bibinfo {author} {\bibfnamefont {L.~S.}\ \bibnamefont
  {Collaboration}},\ }\href@noop {} {\enquote {\bibinfo {title}
  {Ligo/virgo/kagra public alerts},}\ }\bibinfo {howpublished}
  {\url{https://gracedb.ligo.org/superevents/public/O4/}}\BibitemShut {NoStop}%
\bibitem [{\citenamefont {{KAGRA Collaboration}}\ \emph
  {et~al.}(2020)\citenamefont {{KAGRA Collaboration}}, \citenamefont {Akutsu}
  \emph {et~al.}}]{KAGRAscience}%
  \BibitemOpen
  \bibfield  {author} {\bibinfo {author} {\bibnamefont {{KAGRA
  Collaboration}}}, \bibinfo {author} {\bibfnamefont {T.}~\bibnamefont
  {Akutsu}},  \emph {et~al.},\ }\href@noop {} {\enquote {\bibinfo {title}
  {{Overview of KAGRA : KAGRA science}},}\ } (\bibinfo {year} {2020}),\ \Eprint
  {http://arxiv.org/abs/2008.02921} {arXiv:2008.02921 [gr-qc]} \BibitemShut
  {NoStop}%
\bibitem [{\citenamefont {Burtnyk}()}]{LIGO_O4_update}%
  \BibitemOpen
  \bibfield  {author} {\bibinfo {author} {\bibfnamefont {K.}~\bibnamefont
  {Burtnyk}},\ }\href@noop {} {\enquote {\bibinfo {title} {Latest update on
  start of next observing run (o4)},}\ }\bibinfo {howpublished}
  {\url{https://www.ligo.caltech.edu/news/ligo20220123}}\BibitemShut {NoStop}%
\bibitem [{\citenamefont {Messick}\ \emph {et~al.}(2017)\citenamefont
  {Messick}, \citenamefont {Blackburn}, \citenamefont {Brady}, \citenamefont
  {Brockill}, \citenamefont {Cannon}, \citenamefont {Cariou} \emph
  {et~al.}}]{Messick_2017}%
  \BibitemOpen
  \bibfield  {author} {\bibinfo {author} {\bibfnamefont {C.}~\bibnamefont
  {Messick}}, \bibinfo {author} {\bibfnamefont {K.}~\bibnamefont {Blackburn}},
  \bibinfo {author} {\bibfnamefont {P.}~\bibnamefont {Brady}}, \bibinfo
  {author} {\bibfnamefont {P.}~\bibnamefont {Brockill}}, \bibinfo {author}
  {\bibfnamefont {K.}~\bibnamefont {Cannon}}, \bibinfo {author} {\bibfnamefont
  {R.}~\bibnamefont {Cariou}},  \emph {et~al.},\ }\href {\doibase
  10.1103/physrevd.95.042001} {\bibfield  {journal} {\bibinfo  {journal}
  {Physical Review D}\ }\textbf {\bibinfo {volume} {95}},\ \bibinfo {pages}
  {042001} (\bibinfo {year} {2017})}\BibitemShut {NoStop}%
\bibitem [{\citenamefont {Cannon}\ \emph {et~al.}(2021)\citenamefont {Cannon}
  \emph {et~al.}}]{cannon2021gstlal}%
  \BibitemOpen
  \bibfield  {author} {\bibinfo {author} {\bibfnamefont {K.}~\bibnamefont
  {Cannon}} \emph {et~al.},\ }\href {\doibase
  https://doi.org/10.1016/j.softx.2021.100680} {\bibfield  {journal} {\bibinfo
  {journal} {SoftwareX}\ }\textbf {\bibinfo {volume} {14}},\ \bibinfo {pages}
  {100680} (\bibinfo {year} {2021})},\ \Eprint
  {http://arxiv.org/abs/2010.05082} {arXiv:2010.05082 [astro-ph.IM]}
  \BibitemShut {NoStop}%
\bibitem [{\citenamefont {Sachdev}\ \emph {et~al.}(2019)\citenamefont {Sachdev}
  \emph {et~al.}}]{Sachdev2019vvd}%
  \BibitemOpen
  \bibfield  {author} {\bibinfo {author} {\bibfnamefont {S.}~\bibnamefont
  {Sachdev}} \emph {et~al.},\ }\href@noop {} {\  (\bibinfo {year} {2019})},\
  \Eprint {http://arxiv.org/abs/1901.08580} {arXiv:1901.08580 [gr-qc]}
  \BibitemShut {NoStop}%
\bibitem [{\citenamefont {Hanna}\ \emph {et~al.}(2020)\citenamefont {Hanna},
  \citenamefont {Caudill}, \citenamefont {Messick}, \citenamefont {Reza},
  \citenamefont {Sachdev}, \citenamefont {Tsukada} \emph
  {et~al.}}]{hanna_2020}%
  \BibitemOpen
  \bibfield  {author} {\bibinfo {author} {\bibfnamefont {C.}~\bibnamefont
  {Hanna}}, \bibinfo {author} {\bibfnamefont {S.}~\bibnamefont {Caudill}},
  \bibinfo {author} {\bibfnamefont {C.}~\bibnamefont {Messick}}, \bibinfo
  {author} {\bibfnamefont {A.}~\bibnamefont {Reza}}, \bibinfo {author}
  {\bibfnamefont {S.}~\bibnamefont {Sachdev}}, \bibinfo {author} {\bibfnamefont
  {L.}~\bibnamefont {Tsukada}},  \emph {et~al.},\ }\href {\doibase
  10.1103/physrevd.101.022003} {\bibfield  {journal} {\bibinfo  {journal}
  {Physical Review D}\ }\textbf {\bibinfo {volume} {101}},\ \bibinfo {pages}
  {022003} (\bibinfo {year} {2020})}\BibitemShut {NoStop}%
\bibitem [{\citenamefont {{LIGO Scientific Collaboration and Virgo
  Collaboration}}(2023)}]{gstlal}%
  \BibitemOpen
  \bibfield  {author} {\bibinfo {author} {\bibnamefont {{LIGO Scientific
  Collaboration and Virgo Collaboration}}},\ }\href@noop {} {\enquote {\bibinfo
  {title} {Gstlal},}\ }\bibinfo {howpublished}
  {\url{git.ligo.org/lscsoft/gstlal}} (\bibinfo {year} {2023})\BibitemShut
  {NoStop}%
\bibitem [{\citenamefont {Usman}\ \emph {et~al.}(2016)\citenamefont {Usman}
  \emph {et~al.}}]{Usman:2015kfa}%
  \BibitemOpen
  \bibfield  {author} {\bibinfo {author} {\bibfnamefont {S.~A.}\ \bibnamefont
  {Usman}} \emph {et~al.},\ }\href {\doibase 10.1088/0264-9381/33/21/215004}
  {\bibfield  {journal} {\bibinfo  {journal} {Class. Quant. Grav.}\ }\textbf
  {\bibinfo {volume} {33}},\ \bibinfo {pages} {215004} (\bibinfo {year}
  {2016})},\ \Eprint {http://arxiv.org/abs/1508.02357} {arXiv:1508.02357
  [gr-qc]} \BibitemShut {NoStop}%
\bibitem [{\citenamefont {Nitz}\ \emph {et~al.}(2018)\citenamefont {Nitz},
  \citenamefont {Dal~Canton}, \citenamefont {Davis},\ and\ \citenamefont
  {Reyes}}]{PhysRevD.98.024050}%
  \BibitemOpen
  \bibfield  {author} {\bibinfo {author} {\bibfnamefont {A.~H.}\ \bibnamefont
  {Nitz}}, \bibinfo {author} {\bibfnamefont {T.}~\bibnamefont {Dal~Canton}},
  \bibinfo {author} {\bibfnamefont {D.}~\bibnamefont {Davis}}, \ and\ \bibinfo
  {author} {\bibfnamefont {S.}~\bibnamefont {Reyes}},\ }\href {\doibase
  10.1103/PhysRevD.98.024050} {\bibfield  {journal} {\bibinfo  {journal} {Phys.
  Rev. D}\ }\textbf {\bibinfo {volume} {98}},\ \bibinfo {pages} {024050}
  (\bibinfo {year} {2018})}\BibitemShut {NoStop}%
\bibitem [{\citenamefont {Canton}\ \emph {et~al.}(2021)\citenamefont {Canton},
  \citenamefont {Nitz}, \citenamefont {Gadre}, \citenamefont {Davies},
  \citenamefont {Villa-Ortega}, \citenamefont {Dent}, \citenamefont {Harry},\
  and\ \citenamefont {Xiao}}]{Dal_Canton_2021}%
  \BibitemOpen
  \bibfield  {author} {\bibinfo {author} {\bibfnamefont {T.~D.}\ \bibnamefont
  {Canton}}, \bibinfo {author} {\bibfnamefont {A.~H.}\ \bibnamefont {Nitz}},
  \bibinfo {author} {\bibfnamefont {B.}~\bibnamefont {Gadre}}, \bibinfo
  {author} {\bibfnamefont {G.~S.~C.}\ \bibnamefont {Davies}}, \bibinfo {author}
  {\bibfnamefont {V.}~\bibnamefont {Villa-Ortega}}, \bibinfo {author}
  {\bibfnamefont {T.}~\bibnamefont {Dent}}, \bibinfo {author} {\bibfnamefont
  {I.}~\bibnamefont {Harry}}, \ and\ \bibinfo {author} {\bibfnamefont
  {L.}~\bibnamefont {Xiao}},\ }\href {\doibase 10.3847/1538-4357/ac2f9a}
  {\bibfield  {journal} {\bibinfo  {journal} {The Astrophysical Journal}\
  }\textbf {\bibinfo {volume} {923}},\ \bibinfo {pages} {254} (\bibinfo {year}
  {2021})}\BibitemShut {NoStop}%
\bibitem [{\citenamefont {Adams}\ \emph {et~al.}(2016)\citenamefont {Adams},
  \citenamefont {Buskulic}, \citenamefont {Germain}, \citenamefont {Guidi},
  \citenamefont {Marion}, \citenamefont {Montani}, \citenamefont {Mours},
  \citenamefont {Piergiovanni},\ and\ \citenamefont {Wang}}]{Adams:2015ulm}%
  \BibitemOpen
  \bibfield  {author} {\bibinfo {author} {\bibfnamefont {T.}~\bibnamefont
  {Adams}}, \bibinfo {author} {\bibfnamefont {D.}~\bibnamefont {Buskulic}},
  \bibinfo {author} {\bibfnamefont {V.}~\bibnamefont {Germain}}, \bibinfo
  {author} {\bibfnamefont {G.~M.}\ \bibnamefont {Guidi}}, \bibinfo {author}
  {\bibfnamefont {F.}~\bibnamefont {Marion}}, \bibinfo {author} {\bibfnamefont
  {M.}~\bibnamefont {Montani}}, \bibinfo {author} {\bibfnamefont
  {B.}~\bibnamefont {Mours}}, \bibinfo {author} {\bibfnamefont
  {F.}~\bibnamefont {Piergiovanni}}, \ and\ \bibinfo {author} {\bibfnamefont
  {G.}~\bibnamefont {Wang}},\ }\href {\doibase 10.1088/0264-9381/33/17/175012}
  {\bibfield  {journal} {\bibinfo  {journal} {Class. Quant. Grav.}\ }\textbf
  {\bibinfo {volume} {33}},\ \bibinfo {pages} {175012} (\bibinfo {year}
  {2016})},\ \Eprint {http://arxiv.org/abs/1512.02864} {arXiv:1512.02864
  [gr-qc]} \BibitemShut {NoStop}%
\bibitem [{\citenamefont {Aubin}\ \emph {et~al.}(2021)\citenamefont {Aubin}
  \emph {et~al.}}]{Aubin_2021}%
  \BibitemOpen
  \bibfield  {author} {\bibinfo {author} {\bibfnamefont {F.}~\bibnamefont
  {Aubin}} \emph {et~al.},\ }\href {\doibase 10.1088/1361-6382/abe913}
  {\bibfield  {journal} {\bibinfo  {journal} {Classical and Quantum Gravity}\
  }\textbf {\bibinfo {volume} {38}},\ \bibinfo {pages} {095004} (\bibinfo
  {year} {2021})}\BibitemShut {NoStop}%
\bibitem [{\citenamefont {Andres}\ \emph {et~al.}(2022)\citenamefont {Andres},
  \citenamefont {Assiduo}, \citenamefont {Aubin}, \citenamefont {Chierici},
  \citenamefont {Estevez}, \citenamefont {Faedi}, \citenamefont {Guidi},
  \citenamefont {Juste}, \citenamefont {Marion}, \citenamefont {Mours},
  \citenamefont {Nitoglia},\ and\ \citenamefont {Sordini}}]{Andres_2022}%
  \BibitemOpen
  \bibfield  {author} {\bibinfo {author} {\bibfnamefont {N.}~\bibnamefont
  {Andres}}, \bibinfo {author} {\bibfnamefont {M.}~\bibnamefont {Assiduo}},
  \bibinfo {author} {\bibfnamefont {F.}~\bibnamefont {Aubin}}, \bibinfo
  {author} {\bibfnamefont {R.}~\bibnamefont {Chierici}}, \bibinfo {author}
  {\bibfnamefont {D.}~\bibnamefont {Estevez}}, \bibinfo {author} {\bibfnamefont
  {F.}~\bibnamefont {Faedi}}, \bibinfo {author} {\bibfnamefont {G.~M.}\
  \bibnamefont {Guidi}}, \bibinfo {author} {\bibfnamefont {V.}~\bibnamefont
  {Juste}}, \bibinfo {author} {\bibfnamefont {F.}~\bibnamefont {Marion}},
  \bibinfo {author} {\bibfnamefont {B.}~\bibnamefont {Mours}}, \bibinfo
  {author} {\bibfnamefont {E.}~\bibnamefont {Nitoglia}}, \ and\ \bibinfo
  {author} {\bibfnamefont {V.}~\bibnamefont {Sordini}},\ }\href {\doibase
  10.1088/1361-6382/ac482a} {\bibfield  {journal} {\bibinfo  {journal}
  {Classical and Quantum Gravity}\ }\textbf {\bibinfo {volume} {39}},\ \bibinfo
  {pages} {055002} (\bibinfo {year} {2022})}\BibitemShut {NoStop}%
\bibitem [{\citenamefont {Chu}\ \emph {et~al.}(2022)\citenamefont {Chu},
  \citenamefont {Kovalam}, \citenamefont {Wen}, \citenamefont {Slaven-Blair},
  \citenamefont {Bosveld}, \citenamefont {Chen} \emph
  {et~al.}}]{PhysRevD.105.024023}%
  \BibitemOpen
  \bibfield  {author} {\bibinfo {author} {\bibfnamefont {Q.}~\bibnamefont
  {Chu}}, \bibinfo {author} {\bibfnamefont {M.}~\bibnamefont {Kovalam}},
  \bibinfo {author} {\bibfnamefont {L.}~\bibnamefont {Wen}}, \bibinfo {author}
  {\bibfnamefont {T.}~\bibnamefont {Slaven-Blair}}, \bibinfo {author}
  {\bibfnamefont {J.}~\bibnamefont {Bosveld}}, \bibinfo {author} {\bibfnamefont
  {Y.}~\bibnamefont {Chen}},  \emph {et~al.},\ }\href {\doibase
  10.1103/PhysRevD.105.024023} {\bibfield  {journal} {\bibinfo  {journal}
  {Phys. Rev. D}\ }\textbf {\bibinfo {volume} {105}},\ \bibinfo {pages}
  {024023} (\bibinfo {year} {2022})},\ \Eprint
  {http://arxiv.org/abs/2011.06787} {arXiv:2011.06787 [gr-qc]} \BibitemShut
  {NoStop}%
\bibitem [{\citenamefont {Hooper}\ \emph {et~al.}(2012)\citenamefont {Hooper},
  \citenamefont {Chung}, \citenamefont {Luan}, \citenamefont {Blair},
  \citenamefont {Chen},\ and\ \citenamefont {Wen}}]{PhysRevD.86.024012}%
  \BibitemOpen
  \bibfield  {author} {\bibinfo {author} {\bibfnamefont {S.}~\bibnamefont
  {Hooper}}, \bibinfo {author} {\bibfnamefont {S.~K.}\ \bibnamefont {Chung}},
  \bibinfo {author} {\bibfnamefont {J.}~\bibnamefont {Luan}}, \bibinfo {author}
  {\bibfnamefont {D.}~\bibnamefont {Blair}}, \bibinfo {author} {\bibfnamefont
  {Y.}~\bibnamefont {Chen}}, \ and\ \bibinfo {author} {\bibfnamefont
  {L.}~\bibnamefont {Wen}},\ }\href {\doibase 10.1103/PhysRevD.86.024012}
  {\bibfield  {journal} {\bibinfo  {journal} {Phys. Rev. D}\ }\textbf {\bibinfo
  {volume} {86}},\ \bibinfo {pages} {024012} (\bibinfo {year}
  {2012})}\BibitemShut {NoStop}%
\bibitem [{\citenamefont {Chu}(2017)}]{Chu_Thesis}%
  \BibitemOpen
  \bibfield  {author} {\bibinfo {author} {\bibfnamefont {Q.}~\bibnamefont
  {Chu}},\ }\href {\doibase 10.4225/23/5987feb0a789c} {\enquote {\bibinfo
  {title} {Low-latency detection and localization of gravitational waves from
  compact binary coalescences},}\ } (\bibinfo {year} {2017})\BibitemShut
  {NoStop}%
\bibitem [{\citenamefont {Tsukada}\ \emph {et~al.}(2023)\citenamefont {Tsukada}
  \emph {et~al.}}]{Tsukada_LR_2023}%
  \BibitemOpen
  \bibfield  {author} {\bibinfo {author} {\bibfnamefont {L.}~\bibnamefont
  {Tsukada}} \emph {et~al.},\ }\href {\doibase 10.1103/PhysRevD.108.043004}
  {\bibfield  {journal} {\bibinfo  {journal} {Phys. Rev. D}\ }\textbf {\bibinfo
  {volume} {108}},\ \bibinfo {pages} {043004} (\bibinfo {year}
  {2023})}\BibitemShut {NoStop}%
\bibitem [{\citenamefont {Ewing}\ \emph {et~al.}(2023)\citenamefont {Ewing}
  \emph {et~al.}}]{Ewing_performance_2023}%
  \BibitemOpen
  \bibfield  {author} {\bibinfo {author} {\bibfnamefont {B.}~\bibnamefont
  {Ewing}} \emph {et~al.},\ }\href@noop {} {\enquote {\bibinfo {title}
  {Performance of the low-latency gstlal inspiral search towards ligo, virgo,
  and kagra's fourth observing run},}\ } (\bibinfo {year} {2023}),\ \Eprint
  {http://arxiv.org/abs/2305.05625} {arXiv:2305.05625 [gr-qc]} \BibitemShut
  {NoStop}%
\bibitem [{\citenamefont {Sathyaprakash}\ and\ \citenamefont
  {Dhurandhar}(1991)}]{PhysRevD.44.3819}%
  \BibitemOpen
  \bibfield  {author} {\bibinfo {author} {\bibfnamefont {B.~S.}\ \bibnamefont
  {Sathyaprakash}}\ and\ \bibinfo {author} {\bibfnamefont {S.~V.}\ \bibnamefont
  {Dhurandhar}},\ }\href {\doibase 10.1103/PhysRevD.44.3819} {\bibfield
  {journal} {\bibinfo  {journal} {Phys. Rev. D}\ }\textbf {\bibinfo {volume}
  {44}},\ \bibinfo {pages} {3819} (\bibinfo {year} {1991})}\BibitemShut
  {NoStop}%
\bibitem [{\citenamefont {Dhurandhar}\ and\ \citenamefont
  {Sathyaprakash}(1994)}]{PhysRevD.49.1707}%
  \BibitemOpen
  \bibfield  {author} {\bibinfo {author} {\bibfnamefont {S.~V.}\ \bibnamefont
  {Dhurandhar}}\ and\ \bibinfo {author} {\bibfnamefont {B.~S.}\ \bibnamefont
  {Sathyaprakash}},\ }\href {\doibase 10.1103/PhysRevD.49.1707} {\bibfield
  {journal} {\bibinfo  {journal} {Phys. Rev. D}\ }\textbf {\bibinfo {volume}
  {49}},\ \bibinfo {pages} {1707} (\bibinfo {year} {1994})}\BibitemShut
  {NoStop}%
\bibitem [{\citenamefont {Owen}(1996)}]{PhysRevD.53.6749}%
  \BibitemOpen
  \bibfield  {author} {\bibinfo {author} {\bibfnamefont {B.~J.}\ \bibnamefont
  {Owen}},\ }\href {\doibase 10.1103/PhysRevD.53.6749} {\bibfield  {journal}
  {\bibinfo  {journal} {Phys. Rev. D}\ }\textbf {\bibinfo {volume} {53}},\
  \bibinfo {pages} {6749} (\bibinfo {year} {1996})}\BibitemShut {NoStop}%
\bibitem [{\citenamefont {Owen}\ and\ \citenamefont
  {Sathyaprakash}(1999)}]{Owen:1998dk}%
  \BibitemOpen
  \bibfield  {author} {\bibinfo {author} {\bibfnamefont {B.~J.}\ \bibnamefont
  {Owen}}\ and\ \bibinfo {author} {\bibfnamefont {B.~S.}\ \bibnamefont
  {Sathyaprakash}},\ }\href {\doibase 10.1103/PhysRevD.60.022002} {\bibfield
  {journal} {\bibinfo  {journal} {Phys. Rev.}\ }\textbf {\bibinfo {volume}
  {D60}},\ \bibinfo {pages} {022002} (\bibinfo {year} {1999})},\ \Eprint
  {http://arxiv.org/abs/gr-qc/9808076} {arXiv:gr-qc/9808076 [gr-qc]}
  \BibitemShut {NoStop}%
\bibitem [{\citenamefont {Abbott}\ \emph
  {et~al.}(2021{\natexlab{b}})\citenamefont {Abbott} \emph
  {et~al.}}]{abbott2021gwtc}%
  \BibitemOpen
  \bibfield  {author} {\bibinfo {author} {\bibfnamefont {R.}~\bibnamefont
  {Abbott}} \emph {et~al.} (\bibinfo {collaboration} {LIGO Scientific
  Collaboration and Virgo Collaboration}),\ }\href {\doibase
  10.1103/PhysRevX.11.021053} {\bibfield  {journal} {\bibinfo  {journal} {Phys.
  Rev. X}\ }\textbf {\bibinfo {volume} {11}},\ \bibinfo {pages} {021053}
  (\bibinfo {year} {2021}{\natexlab{b}})}\BibitemShut {NoStop}%
\bibitem [{gst(2023)}]{gstreamer}%
  \BibitemOpen
  \href@noop {} {\enquote {\bibinfo {title} {{GStreamer}},}\ }\bibinfo
  {howpublished} {\url{https://gstreamer.freedesktop.org/}} (\bibinfo {year}
  {2023})\BibitemShut {NoStop}%
\bibitem [{\citenamefont {{LIGO Scientific Collaboration}}(2018)}]{lalsuite}%
  \BibitemOpen
  \bibfield  {author} {\bibinfo {author} {\bibnamefont {{LIGO Scientific
  Collaboration}}},\ }\href {\doibase 10.7935/GT1W-FZ16} {\enquote {\bibinfo
  {title} {{LIGO} {A}lgorithm {L}ibrary - {LALS}uite},}\ }\bibinfo
  {howpublished} {free software (GPL)} (\bibinfo {year} {2018})\BibitemShut
  {NoStop}%
\bibitem [{\citenamefont {Abbott}\ \emph
  {et~al.}(2019{\natexlab{e}})\citenamefont {Abbott} \emph
  {et~al.}}]{Abbott_2019}%
  \BibitemOpen
  \bibfield  {author} {\bibinfo {author} {\bibfnamefont {B.~P.}\ \bibnamefont
  {Abbott}} \emph {et~al.},\ }\href {\doibase 10.3847/1538-4357/ab0e8f}
  {\bibfield  {journal} {\bibinfo  {journal} {The Astrophysical Journal}\
  }\textbf {\bibinfo {volume} {875}},\ \bibinfo {pages} {161} (\bibinfo {year}
  {2019}{\natexlab{e}})}\BibitemShut {NoStop}%
\bibitem [{\citenamefont {Mukherjee}\ \emph {et~al.}(2021)\citenamefont
  {Mukherjee}, \citenamefont {Caudill}, \citenamefont {Magee}, \citenamefont
  {Messick}, \citenamefont {Privitera}, \citenamefont {Sachdev} \emph
  {et~al.}}]{Mukherjee_2021}%
  \BibitemOpen
  \bibfield  {author} {\bibinfo {author} {\bibfnamefont {D.}~\bibnamefont
  {Mukherjee}}, \bibinfo {author} {\bibfnamefont {S.}~\bibnamefont {Caudill}},
  \bibinfo {author} {\bibfnamefont {R.}~\bibnamefont {Magee}}, \bibinfo
  {author} {\bibfnamefont {C.}~\bibnamefont {Messick}}, \bibinfo {author}
  {\bibfnamefont {S.}~\bibnamefont {Privitera}}, \bibinfo {author}
  {\bibfnamefont {S.}~\bibnamefont {Sachdev}},  \emph {et~al.},\ }\href
  {\doibase 10.1103/physrevd.103.084047} {\bibfield  {journal} {\bibinfo
  {journal} {Physical Review D}\ }\textbf {\bibinfo {volume} {103}},\ \bibinfo
  {pages} {084047} (\bibinfo {year} {2021})}\BibitemShut {NoStop}%
\bibitem [{\citenamefont {Ajith}\ \emph {et~al.}(2014)\citenamefont {Ajith},
  \citenamefont {Fotopoulos}, \citenamefont {Privitera}, \citenamefont
  {Neunzert}, \citenamefont {Mazumder},\ and\ \citenamefont
  {Weinstein}}]{Ajith_2014}%
  \BibitemOpen
  \bibfield  {author} {\bibinfo {author} {\bibfnamefont {P.}~\bibnamefont
  {Ajith}}, \bibinfo {author} {\bibfnamefont {N.}~\bibnamefont {Fotopoulos}},
  \bibinfo {author} {\bibfnamefont {S.}~\bibnamefont {Privitera}}, \bibinfo
  {author} {\bibfnamefont {A.}~\bibnamefont {Neunzert}}, \bibinfo {author}
  {\bibfnamefont {N.}~\bibnamefont {Mazumder}}, \ and\ \bibinfo {author}
  {\bibfnamefont {A.}~\bibnamefont {Weinstein}},\ }\href {\doibase
  10.1103/physrevd.89.084041} {\bibfield  {journal} {\bibinfo  {journal}
  {Physical Review D}\ }\textbf {\bibinfo {volume} {89}},\ \bibinfo {pages}
  {084041} (\bibinfo {year} {2014})}\BibitemShut {NoStop}%
\bibitem [{\citenamefont {Harry}\ \emph {et~al.}(2014)\citenamefont {Harry},
  \citenamefont {Nitz}, \citenamefont {Brown}, \citenamefont {Lundgren},
  \citenamefont {Ochsner},\ and\ \citenamefont {Keppel}}]{Harry_2014}%
  \BibitemOpen
  \bibfield  {author} {\bibinfo {author} {\bibfnamefont {I.~W.}\ \bibnamefont
  {Harry}}, \bibinfo {author} {\bibfnamefont {A.~H.}\ \bibnamefont {Nitz}},
  \bibinfo {author} {\bibfnamefont {D.~A.}\ \bibnamefont {Brown}}, \bibinfo
  {author} {\bibfnamefont {A.~P.}\ \bibnamefont {Lundgren}}, \bibinfo {author}
  {\bibfnamefont {E.}~\bibnamefont {Ochsner}}, \ and\ \bibinfo {author}
  {\bibfnamefont {D.}~\bibnamefont {Keppel}},\ }\href {\doibase
  10.1103/physrevd.89.024010} {\bibfield  {journal} {\bibinfo  {journal}
  {Physical Review D}\ }\textbf {\bibinfo {volume} {89}},\ \bibinfo {pages}
  {024010} (\bibinfo {year} {2014})}\BibitemShut {NoStop}%
\bibitem [{\citenamefont {Allen}\ \emph {et~al.}(2012)\citenamefont {Allen},
  \citenamefont {Anderson}, \citenamefont {Brady}, \citenamefont {Brown},\ and\
  \citenamefont {Creighton}}]{Allen_2012}%
  \BibitemOpen
  \bibfield  {author} {\bibinfo {author} {\bibfnamefont {B.}~\bibnamefont
  {Allen}}, \bibinfo {author} {\bibfnamefont {W.~G.}\ \bibnamefont {Anderson}},
  \bibinfo {author} {\bibfnamefont {P.~R.}\ \bibnamefont {Brady}}, \bibinfo
  {author} {\bibfnamefont {D.~A.}\ \bibnamefont {Brown}}, \ and\ \bibinfo
  {author} {\bibfnamefont {J.~D.~E.}\ \bibnamefont {Creighton}},\ }\href
  {\doibase 10.1103/physrevd.85.122006} {\bibfield  {journal} {\bibinfo
  {journal} {Physical Review D}\ }\textbf {\bibinfo {volume} {85}} (\bibinfo
  {year} {2012}),\ 10.1103/physrevd.85.122006}\BibitemShut {NoStop}%
\bibitem [{\citenamefont {Privitera}\ \emph {et~al.}(2014)\citenamefont
  {Privitera}, \citenamefont {Mohapatra}, \citenamefont {Ajith}, \citenamefont
  {Cannon}, \citenamefont {Fotopoulos}, \citenamefont {Frei}, \citenamefont
  {Hanna}, \citenamefont {Weinstein},\ and\ \citenamefont
  {Whelan}}]{PhysRevD.89.024003}%
  \BibitemOpen
  \bibfield  {author} {\bibinfo {author} {\bibfnamefont {S.}~\bibnamefont
  {Privitera}}, \bibinfo {author} {\bibfnamefont {S.~R.~P.}\ \bibnamefont
  {Mohapatra}}, \bibinfo {author} {\bibfnamefont {P.}~\bibnamefont {Ajith}},
  \bibinfo {author} {\bibfnamefont {K.}~\bibnamefont {Cannon}}, \bibinfo
  {author} {\bibfnamefont {N.}~\bibnamefont {Fotopoulos}}, \bibinfo {author}
  {\bibfnamefont {M.~A.}\ \bibnamefont {Frei}}, \bibinfo {author}
  {\bibfnamefont {C.}~\bibnamefont {Hanna}}, \bibinfo {author} {\bibfnamefont
  {A.~J.}\ \bibnamefont {Weinstein}}, \ and\ \bibinfo {author} {\bibfnamefont
  {J.~T.}\ \bibnamefont {Whelan}},\ }\href {\doibase
  10.1103/PhysRevD.89.024003} {\bibfield  {journal} {\bibinfo  {journal} {Phys.
  Rev. D}\ }\textbf {\bibinfo {volume} {89}},\ \bibinfo {pages} {024003}
  (\bibinfo {year} {2014})}\BibitemShut {NoStop}%
\bibitem [{\citenamefont {Cannon}\ \emph {et~al.}(2015)\citenamefont {Cannon},
  \citenamefont {Hanna},\ and\ \citenamefont {Peoples}}]{Cannon2015gha}%
  \BibitemOpen
  \bibfield  {author} {\bibinfo {author} {\bibfnamefont {K.}~\bibnamefont
  {Cannon}}, \bibinfo {author} {\bibfnamefont {C.}~\bibnamefont {Hanna}}, \
  and\ \bibinfo {author} {\bibfnamefont {J.}~\bibnamefont {Peoples}},\
  }\href@noop {} {\  (\bibinfo {year} {2015})},\ \Eprint
  {http://arxiv.org/abs/1504.04632} {arXiv:1504.04632 [astro-ph.IM]}
  \BibitemShut {NoStop}%
\bibitem [{\citenamefont {Fong}(2018)}]{Fong2018thesis}%
  \BibitemOpen
  \bibfield  {author} {\bibinfo {author} {\bibfnamefont {H.}~\bibnamefont
  {Fong}},\ }\href {https://tspace.library.utoronto.ca/handle/1807/91831}
  {\enquote {\bibinfo {title} {From simulations to signals: Analyzing
  gravitational waves from compact binary coalescences},}\ } (\bibinfo {year}
  {2018})\BibitemShut {NoStop}%
\bibitem [{\citenamefont {Veitch}\ \emph {et~al.}(2015)\citenamefont {Veitch},
  \citenamefont {Raymond}, \citenamefont {Farr}, \citenamefont {Farr},
  \citenamefont {Graff}, \citenamefont {Vitale} \emph {et~al.}}]{Veitch_2015}%
  \BibitemOpen
  \bibfield  {author} {\bibinfo {author} {\bibfnamefont {J.}~\bibnamefont
  {Veitch}}, \bibinfo {author} {\bibfnamefont {V.}~\bibnamefont {Raymond}},
  \bibinfo {author} {\bibfnamefont {B.}~\bibnamefont {Farr}}, \bibinfo {author}
  {\bibfnamefont {W.}~\bibnamefont {Farr}}, \bibinfo {author} {\bibfnamefont
  {P.}~\bibnamefont {Graff}}, \bibinfo {author} {\bibfnamefont
  {S.}~\bibnamefont {Vitale}},  \emph {et~al.},\ }\href {\doibase
  10.1103/physrevd.91.042003} {\bibfield  {journal} {\bibinfo  {journal}
  {Physical Review D}\ }\textbf {\bibinfo {volume} {91}},\ \bibinfo {pages}
  {042003} (\bibinfo {year} {2015})}\BibitemShut {NoStop}%
\bibitem [{\citenamefont {Ashton}\ \emph {et~al.}(2019)\citenamefont {Ashton}
  \emph {et~al.}}]{Ashton_2019}%
  \BibitemOpen
  \bibfield  {author} {\bibinfo {author} {\bibfnamefont {G.}~\bibnamefont
  {Ashton}} \emph {et~al.},\ }\href {\doibase 10.3847/1538-4365/ab06fc}
  {\bibfield  {journal} {\bibinfo  {journal} {The Astrophysical Journal
  Supplement Series}\ }\textbf {\bibinfo {volume} {241}},\ \bibinfo {pages}
  {27} (\bibinfo {year} {2019})}\BibitemShut {NoStop}%
\bibitem [{\citenamefont {Romero-Shaw}\ \emph {et~al.}(2020)\citenamefont
  {Romero-Shaw} \emph {et~al.}}]{Romero_Shaw_2020}%
  \BibitemOpen
  \bibfield  {author} {\bibinfo {author} {\bibfnamefont {I.~M.}\ \bibnamefont
  {Romero-Shaw}} \emph {et~al.},\ }\href {\doibase 10.1093/mnras/staa2850}
  {\bibfield  {journal} {\bibinfo  {journal} {Monthly Notices of the Royal
  Astronomical Society}\ }\textbf {\bibinfo {volume} {499}},\ \bibinfo {pages}
  {3295} (\bibinfo {year} {2020})},\ \Eprint
  {http://arxiv.org/abs/https://doi.org/10.1093\%2Fmnras\%2Fstaa2850}
  {https://doi.org/10.1093\%2Fmnras\%2Fstaa2850} \BibitemShut {NoStop}%
\bibitem [{\citenamefont {Speagle}(2020)}]{10.1093/mnras/staa278}%
  \BibitemOpen
  \bibfield  {author} {\bibinfo {author} {\bibfnamefont {J.~S.}\ \bibnamefont
  {Speagle}},\ }\href {\doibase 10.1093/mnras/staa278} {\bibfield  {journal}
  {\bibinfo  {journal} {Monthly Notices of the Royal Astronomical Society}\
  }\textbf {\bibinfo {volume} {493}},\ \bibinfo {pages} {3132} (\bibinfo {year}
  {2020})},\ \Eprint
  {http://arxiv.org/abs/https://doi.org/10.1093\%2Fmnras\%2Fstaa278}
  {https://doi.org/10.1093\%2Fmnras\%2Fstaa278} \BibitemShut {NoStop}%
\bibitem [{\citenamefont {Pankow}\ \emph {et~al.}(2015)\citenamefont {Pankow},
  \citenamefont {Brady}, \citenamefont {Ochsner},\ and\ \citenamefont
  {O'Shaughnessy}}]{PhysRevD.92.023002}%
  \BibitemOpen
  \bibfield  {author} {\bibinfo {author} {\bibfnamefont {C.}~\bibnamefont
  {Pankow}}, \bibinfo {author} {\bibfnamefont {P.}~\bibnamefont {Brady}},
  \bibinfo {author} {\bibfnamefont {E.}~\bibnamefont {Ochsner}}, \ and\
  \bibinfo {author} {\bibfnamefont {R.}~\bibnamefont {O'Shaughnessy}},\ }\href
  {\doibase 10.1103/PhysRevD.92.023002} {\bibfield  {journal} {\bibinfo
  {journal} {Phys. Rev. D}\ }\textbf {\bibinfo {volume} {92}},\ \bibinfo
  {pages} {023002} (\bibinfo {year} {2015})}\BibitemShut {NoStop}%
\bibitem [{\citenamefont {Ray}\ \emph {et~al.}(2023)\citenamefont {Ray} \emph
  {et~al.}}]{Ray_pastro_2023}%
  \BibitemOpen
  \bibfield  {author} {\bibinfo {author} {\bibfnamefont {A.}~\bibnamefont
  {Ray}} \emph {et~al.},\ }\href@noop {} {\enquote {\bibinfo {title} {When to
  point your telescopes: Gravitational wave trigger classification for
  real-time multi-messenger followup observations},}\ } (\bibinfo {year}
  {2023}),\ \Eprint {http://arxiv.org/abs/2306.07190} {arXiv:2306.07190
  [gr-qc]} \BibitemShut {NoStop}%
\bibitem [{\citenamefont {Abbott}\ \emph
  {et~al.}(2016{\natexlab{d}})\citenamefont {Abbott} \emph
  {et~al.}}]{Abbott_2016_150914}%
  \BibitemOpen
  \bibfield  {author} {\bibinfo {author} {\bibfnamefont {B.~P.}\ \bibnamefont
  {Abbott}} \emph {et~al.},\ }\href {\doibase 10.1103/physrevd.93.122003}
  {\bibfield  {journal} {\bibinfo  {journal} {Physical Review D}\ }\textbf
  {\bibinfo {volume} {93}} (\bibinfo {year} {2016}{\natexlab{d}}),\
  10.1103/physrevd.93.122003}\BibitemShut {NoStop}%
\bibitem [{\citenamefont {Capano}\ \emph {et~al.}(2016)\citenamefont {Capano},
  \citenamefont {Harry}, \citenamefont {Privitera},\ and\ \citenamefont
  {Buonanno}}]{Capano_2016}%
  \BibitemOpen
  \bibfield  {author} {\bibinfo {author} {\bibfnamefont {C.}~\bibnamefont
  {Capano}}, \bibinfo {author} {\bibfnamefont {I.}~\bibnamefont {Harry}},
  \bibinfo {author} {\bibfnamefont {S.}~\bibnamefont {Privitera}}, \ and\
  \bibinfo {author} {\bibfnamefont {A.}~\bibnamefont {Buonanno}},\ }\href
  {\doibase 10.1103/physrevd.93.124007} {\bibfield  {journal} {\bibinfo
  {journal} {Physical Review D}\ }\textbf {\bibinfo {volume} {93}},\ \bibinfo
  {pages} {124007} (\bibinfo {year} {2016})}\BibitemShut {NoStop}%
\bibitem [{\citenamefont {Harry}\ \emph {et~al.}(2009)\citenamefont {Harry},
  \citenamefont {Allen},\ and\ \citenamefont {Sathyaprakash}}]{Harry_2009}%
  \BibitemOpen
  \bibfield  {author} {\bibinfo {author} {\bibfnamefont {I.~W.}\ \bibnamefont
  {Harry}}, \bibinfo {author} {\bibfnamefont {B.}~\bibnamefont {Allen}}, \ and\
  \bibinfo {author} {\bibfnamefont {B.~S.}\ \bibnamefont {Sathyaprakash}},\
  }\href {\doibase 10.1103/physrevd.80.104014} {\bibfield  {journal} {\bibinfo
  {journal} {Physical Review D}\ }\textbf {\bibinfo {volume} {80}},\ \bibinfo
  {pages} {104014} (\bibinfo {year} {2009})}\BibitemShut {NoStop}%
\bibitem [{\citenamefont {Hanna}\ \emph {et~al.}(2023)\citenamefont {Hanna}
  \emph {et~al.}}]{MANIFOLD}%
  \BibitemOpen
  \bibfield  {author} {\bibinfo {author} {\bibfnamefont {C.}~\bibnamefont
  {Hanna}} \emph {et~al.},\ }\href {\doibase 10.1103/PhysRevD.108.042003}
  {\bibfield  {journal} {\bibinfo  {journal} {Phys. Rev. D}\ }\textbf {\bibinfo
  {volume} {108}},\ \bibinfo {pages} {042003} (\bibinfo {year}
  {2023})}\BibitemShut {NoStop}%
\bibitem [{lig(2022)}]{ligo_psd_dcc}%
  \BibitemOpen
  \href {https://dcc.ligo.org/public/0165/T2000012/002/aligo_O4high.txt}
  {\enquote {\bibinfo {title} {Ligo sensitivity (190 mpc) (used for o4
  simulations)},}\ } (\bibinfo {year} {2022})\BibitemShut {NoStop}%
\bibitem [{v1_(2022)}]{v1_psd_dcc}%
  \BibitemOpen
  \href {https://dcc.ligo.org/public/0165/T2000012/002/avirgo_O4high_NEW.txt}
  {\enquote {\bibinfo {title} {Virgo sensitivity (used for o4 simulations)},}\
  } (\bibinfo {year} {2022})\BibitemShut {NoStop}%
\bibitem [{k1_(2022)}]{k1_psd_dcc}%
  \BibitemOpen
  \href {https://dcc.ligo.org/public/0165/T2000012/002/kagra_10Mpc.txt}
  {\enquote {\bibinfo {title} {Kagra sensitivity (10 mpc) (used for o4 high
  sensitivity simulations)},}\ } (\bibinfo {year} {2022})\BibitemShut {NoStop}%
\bibitem [{psd(2022)}]{psd_dcc}%
  \BibitemOpen
  \href {https://dcc.ligo.org/LIGO-T2000012/public} {\enquote {\bibinfo {title}
  {Noise curves used for simulations in the update of the observing scenarios
  paper},}\ } (\bibinfo {year} {2022})\BibitemShut {NoStop}%
\bibitem [{\citenamefont {Moore}\ \emph {et~al.}(2014)\citenamefont {Moore},
  \citenamefont {Cole},\ and\ \citenamefont {Berry}}]{Moore_2014}%
  \BibitemOpen
  \bibfield  {author} {\bibinfo {author} {\bibfnamefont {C.~J.}\ \bibnamefont
  {Moore}}, \bibinfo {author} {\bibfnamefont {R.~H.}\ \bibnamefont {Cole}}, \
  and\ \bibinfo {author} {\bibfnamefont {C.~P.~L.}\ \bibnamefont {Berry}},\
  }\href {\doibase 10.1088/0264-9381/32/1/015014} {\bibfield  {journal}
  {\bibinfo  {journal} {Classical and Quantum Gravity}\ }\textbf {\bibinfo
  {volume} {32}},\ \bibinfo {pages} {015014} (\bibinfo {year}
  {2014})}\BibitemShut {NoStop}%
\bibitem [{\citenamefont {Abbott}\ \emph {et~al.}(2020)\citenamefont {Abbott}
  \emph {et~al.}}]{Abbott_2020}%
  \BibitemOpen
  \bibfield  {author} {\bibinfo {author} {\bibfnamefont {R.}~\bibnamefont
  {Abbott}} \emph {et~al.},\ }\href {\doibase 10.3847/2041-8213/ab960f}
  {\bibfield  {journal} {\bibinfo  {journal} {The Astrophysical Journal
  Letters}\ }\textbf {\bibinfo {volume} {896}},\ \bibinfo {pages} {L44}
  (\bibinfo {year} {2020})}\BibitemShut {NoStop}%
\bibitem [{\citenamefont {Legred}\ \emph {et~al.}(2021)\citenamefont {Legred},
  \citenamefont {Chatziioannou}, \citenamefont {Essick}, \citenamefont {Han},\
  and\ \citenamefont {Landry}}]{Legred_2021}%
  \BibitemOpen
  \bibfield  {author} {\bibinfo {author} {\bibfnamefont {I.}~\bibnamefont
  {Legred}}, \bibinfo {author} {\bibfnamefont {K.}~\bibnamefont
  {Chatziioannou}}, \bibinfo {author} {\bibfnamefont {R.}~\bibnamefont
  {Essick}}, \bibinfo {author} {\bibfnamefont {S.}~\bibnamefont {Han}}, \ and\
  \bibinfo {author} {\bibfnamefont {P.}~\bibnamefont {Landry}},\ }\href
  {\doibase 10.1103/physrevd.104.063003} {\bibfield  {journal} {\bibinfo
  {journal} {Physical Review D}\ }\textbf {\bibinfo {volume} {104}},\ \bibinfo
  {pages} {063003} (\bibinfo {year} {2021})}\BibitemShut {NoStop}%
\bibitem [{\citenamefont {Dietrich}\ \emph {et~al.}(2020)\citenamefont
  {Dietrich}, \citenamefont {Coughlin}, \citenamefont {Pang}, \citenamefont
  {Bulla}, \citenamefont {Heinzel}, \citenamefont {Issa}, \citenamefont
  {Tews},\ and\ \citenamefont {Antier}}]{Dietrich_2020}%
  \BibitemOpen
  \bibfield  {author} {\bibinfo {author} {\bibfnamefont {T.}~\bibnamefont
  {Dietrich}}, \bibinfo {author} {\bibfnamefont {M.~W.}\ \bibnamefont
  {Coughlin}}, \bibinfo {author} {\bibfnamefont {P.~T.~H.}\ \bibnamefont
  {Pang}}, \bibinfo {author} {\bibfnamefont {M.}~\bibnamefont {Bulla}},
  \bibinfo {author} {\bibfnamefont {J.}~\bibnamefont {Heinzel}}, \bibinfo
  {author} {\bibfnamefont {L.}~\bibnamefont {Issa}}, \bibinfo {author}
  {\bibfnamefont {I.}~\bibnamefont {Tews}}, \ and\ \bibinfo {author}
  {\bibfnamefont {S.}~\bibnamefont {Antier}},\ }\href {\doibase
  10.1126/science.abb4317} {\bibfield  {journal} {\bibinfo  {journal}
  {Science}\ }\textbf {\bibinfo {volume} {370}},\ \bibinfo {pages} {1450}
  (\bibinfo {year} {2020})},\ \Eprint
  {http://arxiv.org/abs/https://www.science.org/doi/pdf/10.1126/science.abb4317}
  {https://www.science.org/doi/pdf/10.1126/science.abb4317} \BibitemShut
  {NoStop}%
\bibitem [{\citenamefont {Jiang}\ \emph {et~al.}(2020)\citenamefont {Jiang},
  \citenamefont {Tang}, \citenamefont {Wang}, \citenamefont {Fan},\ and\
  \citenamefont {Wei}}]{Jiang_2020}%
  \BibitemOpen
  \bibfield  {author} {\bibinfo {author} {\bibfnamefont {J.-L.}\ \bibnamefont
  {Jiang}}, \bibinfo {author} {\bibfnamefont {S.-P.}\ \bibnamefont {Tang}},
  \bibinfo {author} {\bibfnamefont {Y.-Z.}\ \bibnamefont {Wang}}, \bibinfo
  {author} {\bibfnamefont {Y.-Z.}\ \bibnamefont {Fan}}, \ and\ \bibinfo
  {author} {\bibfnamefont {D.-M.}\ \bibnamefont {Wei}},\ }\href {\doibase
  10.3847/1538-4357/ab77cf} {\bibfield  {journal} {\bibinfo  {journal} {The
  Astrophysical Journal}\ }\textbf {\bibinfo {volume} {892}},\ \bibinfo {pages}
  {55} (\bibinfo {year} {2020})}\BibitemShut {NoStop}%
\bibitem [{\citenamefont {Abbott}\ \emph {et~al.}(2018)\citenamefont {Abbott}
  \emph {et~al.}}]{Abbott_2018}%
  \BibitemOpen
  \bibfield  {author} {\bibinfo {author} {\bibfnamefont {B.~P.}\ \bibnamefont
  {Abbott}} \emph {et~al.} (\bibinfo {collaboration} {LIGO Scientific,
  Virgo}),\ }\href {\doibase 10.1103/physrevlett.121.161101} {\bibfield
  {journal} {\bibinfo  {journal} {Physical Review Letters}\ }\textbf {\bibinfo
  {volume} {121}} (\bibinfo {year} {2018}),\
  10.1103/physrevlett.121.161101}\BibitemShut {NoStop}%
\bibitem [{\citenamefont {Cromartie}\ \emph {et~al.}(2019)\citenamefont
  {Cromartie} \emph {et~al.}}]{Cromartie_2019}%
  \BibitemOpen
  \bibfield  {author} {\bibinfo {author} {\bibfnamefont {H.~T.}\ \bibnamefont
  {Cromartie}} \emph {et~al.},\ }\href {\doibase 10.1038/s41550-019-0880-2}
  {\bibfield  {journal} {\bibinfo  {journal} {Nature Astronomy}\ }\textbf
  {\bibinfo {volume} {4}},\ \bibinfo {pages} {72} (\bibinfo {year}
  {2019})}\BibitemShut {NoStop}%
\bibitem [{\citenamefont {Kalogera}\ and\ \citenamefont
  {Baym}(1996)}]{Kalogera_1996}%
  \BibitemOpen
  \bibfield  {author} {\bibinfo {author} {\bibfnamefont {V.}~\bibnamefont
  {Kalogera}}\ and\ \bibinfo {author} {\bibfnamefont {G.}~\bibnamefont
  {Baym}},\ }\href {\doibase 10.1086/310296} {\bibfield  {journal} {\bibinfo
  {journal} {The Astrophysical Journal}\ }\textbf {\bibinfo {volume} {470}},\
  \bibinfo {pages} {L61} (\bibinfo {year} {1996})}\BibitemShut {NoStop}%
\bibitem [{\citenamefont {Kramer}\ and\ \citenamefont
  {Wex}(2009)}]{Kramer_2009}%
  \BibitemOpen
  \bibfield  {author} {\bibinfo {author} {\bibfnamefont {M.}~\bibnamefont
  {Kramer}}\ and\ \bibinfo {author} {\bibfnamefont {N.}~\bibnamefont {Wex}},\
  }\href {\doibase 10.1088/0264-9381/26/7/073001} {\bibfield  {journal}
  {\bibinfo  {journal} {Classical and Quantum Gravity}\ }\textbf {\bibinfo
  {volume} {26}},\ \bibinfo {pages} {073001} (\bibinfo {year}
  {2009})}\BibitemShut {NoStop}%
\bibitem [{\citenamefont {Burgay}\ \emph {et~al.}(2003)\citenamefont {Burgay}
  \emph {et~al.}}]{Burgay_2003}%
  \BibitemOpen
  \bibfield  {author} {\bibinfo {author} {\bibfnamefont {M.}~\bibnamefont
  {Burgay}} \emph {et~al.},\ }\href {\doibase 10.1038/nature02124} {\bibfield
  {journal} {\bibinfo  {journal} {Nature}\ }\textbf {\bibinfo {volume} {426}},\
  \bibinfo {pages} {531} (\bibinfo {year} {2003})}\BibitemShut {NoStop}%
\bibitem [{\citenamefont {Stovall}\ \emph {et~al.}(2018)\citenamefont {Stovall}
  \emph {et~al.}}]{Stovall_2018}%
  \BibitemOpen
  \bibfield  {author} {\bibinfo {author} {\bibfnamefont {K.}~\bibnamefont
  {Stovall}} \emph {et~al.},\ }\href {\doibase 10.3847/2041-8213/aaad06}
  {\bibfield  {journal} {\bibinfo  {journal} {The Astrophysical Journal}\
  }\textbf {\bibinfo {volume} {854}},\ \bibinfo {pages} {L22} (\bibinfo {year}
  {2018})}\BibitemShut {NoStop}%
\bibitem [{\citenamefont {Gou}\ \emph {et~al.}(2011)\citenamefont {Gou},
  \citenamefont {McClintock}, \citenamefont {Reid}, \citenamefont {Orosz},
  \citenamefont {Steiner}, \citenamefont {Narayan}, \citenamefont {Xiang},
  \citenamefont {Remillard}, \citenamefont {Arnaud},\ and\ \citenamefont
  {Davis}}]{Gou_2011}%
  \BibitemOpen
  \bibfield  {author} {\bibinfo {author} {\bibfnamefont {L.}~\bibnamefont
  {Gou}}, \bibinfo {author} {\bibfnamefont {J.~E.}\ \bibnamefont {McClintock}},
  \bibinfo {author} {\bibfnamefont {M.~J.}\ \bibnamefont {Reid}}, \bibinfo
  {author} {\bibfnamefont {J.~A.}\ \bibnamefont {Orosz}}, \bibinfo {author}
  {\bibfnamefont {J.~F.}\ \bibnamefont {Steiner}}, \bibinfo {author}
  {\bibfnamefont {R.}~\bibnamefont {Narayan}}, \bibinfo {author} {\bibfnamefont
  {J.}~\bibnamefont {Xiang}}, \bibinfo {author} {\bibfnamefont {R.~A.}\
  \bibnamefont {Remillard}}, \bibinfo {author} {\bibfnamefont {K.~A.}\
  \bibnamefont {Arnaud}}, \ and\ \bibinfo {author} {\bibfnamefont {S.~W.}\
  \bibnamefont {Davis}},\ }\href {\doibase 10.1088/0004-637x/742/2/85}
  {\bibfield  {journal} {\bibinfo  {journal} {The Astrophysical Journal}\
  }\textbf {\bibinfo {volume} {742}},\ \bibinfo {pages} {85} (\bibinfo {year}
  {2011})}\BibitemShut {NoStop}%
\bibitem [{\citenamefont {McClintock}\ \emph {et~al.}(2011)\citenamefont
  {McClintock}, \citenamefont {Narayan}, \citenamefont {Davis}, \citenamefont
  {Gou}, \citenamefont {Kulkarni}, \citenamefont {Orosz}, \citenamefont
  {Penna}, \citenamefont {Remillard},\ and\ \citenamefont
  {Steiner}}]{McClintock_2011}%
  \BibitemOpen
  \bibfield  {author} {\bibinfo {author} {\bibfnamefont {J.~E.}\ \bibnamefont
  {McClintock}}, \bibinfo {author} {\bibfnamefont {R.}~\bibnamefont {Narayan}},
  \bibinfo {author} {\bibfnamefont {S.~W.}\ \bibnamefont {Davis}}, \bibinfo
  {author} {\bibfnamefont {L.}~\bibnamefont {Gou}}, \bibinfo {author}
  {\bibfnamefont {A.}~\bibnamefont {Kulkarni}}, \bibinfo {author}
  {\bibfnamefont {J.~A.}\ \bibnamefont {Orosz}}, \bibinfo {author}
  {\bibfnamefont {R.~F.}\ \bibnamefont {Penna}}, \bibinfo {author}
  {\bibfnamefont {R.~A.}\ \bibnamefont {Remillard}}, \ and\ \bibinfo {author}
  {\bibfnamefont {J.~F.}\ \bibnamefont {Steiner}},\ }\href {\doibase
  10.1088/0264-9381/28/11/114009} {\bibfield  {journal} {\bibinfo  {journal}
  {Classical and Quantum Gravity}\ }\textbf {\bibinfo {volume} {28}},\ \bibinfo
  {pages} {114009} (\bibinfo {year} {2011})}\BibitemShut {NoStop}%
\bibitem [{\citenamefont {Fabian}\ \emph {et~al.}(2012)\citenamefont {Fabian}
  \emph {et~al.}}]{Fabian_2012}%
  \BibitemOpen
  \bibfield  {author} {\bibinfo {author} {\bibfnamefont {A.~C.}\ \bibnamefont
  {Fabian}} \emph {et~al.},\ }\href {\doibase 10.1111/j.1365-2966.2012.21185.x}
  {\bibfield  {journal} {\bibinfo  {journal} {Monthly Notices of the Royal
  Astronomical Society}\ }\textbf {\bibinfo {volume} {424}},\ \bibinfo {pages}
  {217} (\bibinfo {year} {2012})}\BibitemShut {NoStop}%
\bibitem [{\citenamefont {Farr}\ \emph {et~al.}(2011)\citenamefont {Farr},
  \citenamefont {Sravan}, \citenamefont {Cantrell}, \citenamefont {Kreidberg},
  \citenamefont {Bailyn}, \citenamefont {Mandel},\ and\ \citenamefont
  {Kalogera}}]{Farr_2011}%
  \BibitemOpen
  \bibfield  {author} {\bibinfo {author} {\bibfnamefont {W.~M.}\ \bibnamefont
  {Farr}}, \bibinfo {author} {\bibfnamefont {N.}~\bibnamefont {Sravan}},
  \bibinfo {author} {\bibfnamefont {A.}~\bibnamefont {Cantrell}}, \bibinfo
  {author} {\bibfnamefont {L.}~\bibnamefont {Kreidberg}}, \bibinfo {author}
  {\bibfnamefont {C.~D.}\ \bibnamefont {Bailyn}}, \bibinfo {author}
  {\bibfnamefont {I.}~\bibnamefont {Mandel}}, \ and\ \bibinfo {author}
  {\bibfnamefont {V.}~\bibnamefont {Kalogera}},\ }\href {\doibase
  10.1088/0004-637x/741/2/103} {\bibfield  {journal} {\bibinfo  {journal} {The
  Astrophysical Journal}\ }\textbf {\bibinfo {volume} {741}},\ \bibinfo {pages}
  {103} (\bibinfo {year} {2011})}\BibitemShut {NoStop}%
\bibitem [{\citenamefont {Özel}\ \emph {et~al.}(2010)\citenamefont {Özel},
  \citenamefont {Psaltis}, \citenamefont {Narayan},\ and\ \citenamefont
  {McClintock}}]{Ozel_2010}%
  \BibitemOpen
  \bibfield  {author} {\bibinfo {author} {\bibfnamefont {F.}~\bibnamefont
  {Özel}}, \bibinfo {author} {\bibfnamefont {D.}~\bibnamefont {Psaltis}},
  \bibinfo {author} {\bibfnamefont {R.}~\bibnamefont {Narayan}}, \ and\
  \bibinfo {author} {\bibfnamefont {J.~E.}\ \bibnamefont {McClintock}},\ }\href
  {\doibase 10.1088/0004-637x/725/2/1918} {\bibfield  {journal} {\bibinfo
  {journal} {The Astrophysical Journal}\ }\textbf {\bibinfo {volume} {725}},\
  \bibinfo {pages} {1918} (\bibinfo {year} {2010})}\BibitemShut {NoStop}%
\bibitem [{\citenamefont {Farmer}\ \emph {et~al.}(2019)\citenamefont {Farmer},
  \citenamefont {Renzo}, \citenamefont {de~Mink}, \citenamefont {Marchant},\
  and\ \citenamefont {Justham}}]{Farmer_2019}%
  \BibitemOpen
  \bibfield  {author} {\bibinfo {author} {\bibfnamefont {R.}~\bibnamefont
  {Farmer}}, \bibinfo {author} {\bibfnamefont {M.}~\bibnamefont {Renzo}},
  \bibinfo {author} {\bibfnamefont {S.~E.}\ \bibnamefont {de~Mink}}, \bibinfo
  {author} {\bibfnamefont {P.}~\bibnamefont {Marchant}}, \ and\ \bibinfo
  {author} {\bibfnamefont {S.}~\bibnamefont {Justham}},\ }\href {\doibase
  10.3847/1538-4357/ab518b} {\bibfield  {journal} {\bibinfo  {journal} {The
  Astrophysical Journal}\ }\textbf {\bibinfo {volume} {887}},\ \bibinfo {pages}
  {53} (\bibinfo {year} {2019})}\BibitemShut {NoStop}%
\bibitem [{\citenamefont {Heger}\ \emph {et~al.}(2003)\citenamefont {Heger},
  \citenamefont {Fryer}, \citenamefont {Woosley}, \citenamefont {Langer},\ and\
  \citenamefont {Hartmann}}]{Heger_2003}%
  \BibitemOpen
  \bibfield  {author} {\bibinfo {author} {\bibfnamefont {A.}~\bibnamefont
  {Heger}}, \bibinfo {author} {\bibfnamefont {C.~L.}\ \bibnamefont {Fryer}},
  \bibinfo {author} {\bibfnamefont {S.~E.}\ \bibnamefont {Woosley}}, \bibinfo
  {author} {\bibfnamefont {N.}~\bibnamefont {Langer}}, \ and\ \bibinfo {author}
  {\bibfnamefont {D.~H.}\ \bibnamefont {Hartmann}},\ }\href {\doibase
  10.1086/375341} {\bibfield  {journal} {\bibinfo  {journal} {The Astrophysical
  Journal}\ }\textbf {\bibinfo {volume} {591}},\ \bibinfo {pages} {288}
  (\bibinfo {year} {2003})}\BibitemShut {NoStop}%
\bibitem [{\citenamefont {Ajith}\ \emph {et~al.}(2011)\citenamefont {Ajith},
  \citenamefont {Hannam}, \citenamefont {Husa}, \citenamefont {Chen},
  \citenamefont {Br\"ugmann}, \citenamefont {Dorband}, \citenamefont
  {M\"uller}, \citenamefont {Ohme}, \citenamefont {Pollney}, \citenamefont
  {Reisswig}, \citenamefont {Santamar\'{\i}a},\ and\ \citenamefont
  {Seiler}}]{PhysRevLett.106.241101}%
  \BibitemOpen
  \bibfield  {author} {\bibinfo {author} {\bibfnamefont {P.}~\bibnamefont
  {Ajith}}, \bibinfo {author} {\bibfnamefont {M.}~\bibnamefont {Hannam}},
  \bibinfo {author} {\bibfnamefont {S.}~\bibnamefont {Husa}}, \bibinfo {author}
  {\bibfnamefont {Y.}~\bibnamefont {Chen}}, \bibinfo {author} {\bibfnamefont
  {B.}~\bibnamefont {Br\"ugmann}}, \bibinfo {author} {\bibfnamefont
  {N.}~\bibnamefont {Dorband}}, \bibinfo {author} {\bibfnamefont
  {D.}~\bibnamefont {M\"uller}}, \bibinfo {author} {\bibfnamefont
  {F.}~\bibnamefont {Ohme}}, \bibinfo {author} {\bibfnamefont {D.}~\bibnamefont
  {Pollney}}, \bibinfo {author} {\bibfnamefont {C.}~\bibnamefont {Reisswig}},
  \bibinfo {author} {\bibfnamefont {L.}~\bibnamefont {Santamar\'{\i}a}}, \ and\
  \bibinfo {author} {\bibfnamefont {J.}~\bibnamefont {Seiler}},\ }\href
  {\doibase 10.1103/PhysRevLett.106.241101} {\bibfield  {journal} {\bibinfo
  {journal} {Phys. Rev. Lett.}\ }\textbf {\bibinfo {volume} {106}},\ \bibinfo
  {pages} {241101} (\bibinfo {year} {2011})}\BibitemShut {NoStop}%
\bibitem [{\citenamefont {Ossokine}\ \emph {et~al.}(2020)\citenamefont
  {Ossokine} \emph {et~al.}}]{Ossokine_2020}%
  \BibitemOpen
  \bibfield  {author} {\bibinfo {author} {\bibfnamefont {S.}~\bibnamefont
  {Ossokine}} \emph {et~al.},\ }\href {\doibase 10.1103/physrevd.102.044055}
  {\bibfield  {journal} {\bibinfo  {journal} {Physical Review D}\ }\textbf
  {\bibinfo {volume} {102}},\ \bibinfo {pages} {044055} (\bibinfo {year}
  {2020})}\BibitemShut {NoStop}%
\bibitem [{\citenamefont {Khan}\ \emph {et~al.}(2016)\citenamefont {Khan},
  \citenamefont {Husa}, \citenamefont {Hannam}, \citenamefont {Ohme},
  \citenamefont {P\"urrer}, \citenamefont {Forteza},\ and\ \citenamefont
  {Boh\'e}}]{Khan_2016}%
  \BibitemOpen
  \bibfield  {author} {\bibinfo {author} {\bibfnamefont {S.}~\bibnamefont
  {Khan}}, \bibinfo {author} {\bibfnamefont {S.}~\bibnamefont {Husa}}, \bibinfo
  {author} {\bibfnamefont {M.}~\bibnamefont {Hannam}}, \bibinfo {author}
  {\bibfnamefont {F.}~\bibnamefont {Ohme}}, \bibinfo {author} {\bibfnamefont
  {M.}~\bibnamefont {P\"urrer}}, \bibinfo {author} {\bibfnamefont {X.~J.}\
  \bibnamefont {Forteza}}, \ and\ \bibinfo {author} {\bibfnamefont
  {A.}~\bibnamefont {Boh\'e}},\ }\href {\doibase 10.1103/physrevd.93.044007}
  {\bibfield  {journal} {\bibinfo  {journal} {Physical Review D}\ }\textbf
  {\bibinfo {volume} {93}},\ \bibinfo {pages} {044007} (\bibinfo {year}
  {2016})}\BibitemShut {NoStop}%
\bibitem [{\citenamefont {Husa}\ \emph {et~al.}(2016)\citenamefont {Husa},
  \citenamefont {Khan}, \citenamefont {Hannam}, \citenamefont {P\"urrer},
  \citenamefont {Ohme}, \citenamefont {Forteza},\ and\ \citenamefont
  {Boh\'e}}]{Husa_2016}%
  \BibitemOpen
  \bibfield  {author} {\bibinfo {author} {\bibfnamefont {S.}~\bibnamefont
  {Husa}}, \bibinfo {author} {\bibfnamefont {S.}~\bibnamefont {Khan}}, \bibinfo
  {author} {\bibfnamefont {M.}~\bibnamefont {Hannam}}, \bibinfo {author}
  {\bibfnamefont {M.}~\bibnamefont {P\"urrer}}, \bibinfo {author}
  {\bibfnamefont {F.}~\bibnamefont {Ohme}}, \bibinfo {author} {\bibfnamefont
  {X.~J.}\ \bibnamefont {Forteza}}, \ and\ \bibinfo {author} {\bibfnamefont
  {A.}~\bibnamefont {Boh\'e}},\ }\href {\doibase 10.1103/physrevd.93.044006}
  {\bibfield  {journal} {\bibinfo  {journal} {Physical Review D}\ }\textbf
  {\bibinfo {volume} {93}},\ \bibinfo {pages} {044006} (\bibinfo {year}
  {2016})}\BibitemShut {NoStop}%
\bibitem [{\citenamefont {Cannon}\ \emph {et~al.}(2010)\citenamefont {Cannon},
  \citenamefont {Chapman}, \citenamefont {Hanna}, \citenamefont {Keppel},
  \citenamefont {Searle},\ and\ \citenamefont {Weinstein}}]{Cannon_2010}%
  \BibitemOpen
  \bibfield  {author} {\bibinfo {author} {\bibfnamefont {K.}~\bibnamefont
  {Cannon}}, \bibinfo {author} {\bibfnamefont {A.}~\bibnamefont {Chapman}},
  \bibinfo {author} {\bibfnamefont {C.}~\bibnamefont {Hanna}}, \bibinfo
  {author} {\bibfnamefont {D.}~\bibnamefont {Keppel}}, \bibinfo {author}
  {\bibfnamefont {A.~C.}\ \bibnamefont {Searle}}, \ and\ \bibinfo {author}
  {\bibfnamefont {A.~J.}\ \bibnamefont {Weinstein}},\ }\href {\doibase
  10.1103/PhysRevD.82.044025} {\bibfield  {journal} {\bibinfo  {journal} {Phys.
  Rev. D}\ }\textbf {\bibinfo {volume} {82}},\ \bibinfo {pages} {044025}
  (\bibinfo {year} {2010})}\BibitemShut {NoStop}%
\bibitem [{\citenamefont {Cannon}\ \emph {et~al.}(2012)\citenamefont {Cannon}
  \emph {et~al.}}]{Cannon_2012}%
  \BibitemOpen
  \bibfield  {author} {\bibinfo {author} {\bibfnamefont {K.}~\bibnamefont
  {Cannon}} \emph {et~al.},\ }\href {\doibase 10.1088/0004-637X/748/2/136}
  {\bibfield  {journal} {\bibinfo  {journal} {The Astrophysical Journal}\
  }\textbf {\bibinfo {volume} {748}},\ \bibinfo {pages} {136} (\bibinfo {year}
  {2012})}\BibitemShut {NoStop}%
\bibitem [{\citenamefont {Canizares}\ \emph {et~al.}(2015)\citenamefont
  {Canizares}, \citenamefont {Field}, \citenamefont {Gair}, \citenamefont
  {Raymond}, \citenamefont {Smith},\ and\ \citenamefont
  {Tiglio}}]{Canizares:2014fya}%
  \BibitemOpen
  \bibfield  {author} {\bibinfo {author} {\bibfnamefont {P.}~\bibnamefont
  {Canizares}}, \bibinfo {author} {\bibfnamefont {S.~E.}\ \bibnamefont
  {Field}}, \bibinfo {author} {\bibfnamefont {J.}~\bibnamefont {Gair}},
  \bibinfo {author} {\bibfnamefont {V.}~\bibnamefont {Raymond}}, \bibinfo
  {author} {\bibfnamefont {R.}~\bibnamefont {Smith}}, \ and\ \bibinfo {author}
  {\bibfnamefont {M.}~\bibnamefont {Tiglio}},\ }\href {\doibase
  10.1103/PhysRevLett.114.071104} {\bibfield  {journal} {\bibinfo  {journal}
  {Phys. Rev. Lett.}\ }\textbf {\bibinfo {volume} {114}},\ \bibinfo {pages}
  {071104} (\bibinfo {year} {2015})},\ \Eprint {http://arxiv.org/abs/1404.6284}
  {arXiv:1404.6284 [gr-qc]} \BibitemShut {NoStop}%
\bibitem [{\citenamefont {Smith}\ \emph {et~al.}(2016)\citenamefont {Smith},
  \citenamefont {Field}, \citenamefont {Blackburn}, \citenamefont {Haster},
  \citenamefont {P\"urrer}, \citenamefont {Raymond},\ and\ \citenamefont
  {Schmidt}}]{Smith:2016qas}%
  \BibitemOpen
  \bibfield  {author} {\bibinfo {author} {\bibfnamefont {R.}~\bibnamefont
  {Smith}}, \bibinfo {author} {\bibfnamefont {S.~E.}\ \bibnamefont {Field}},
  \bibinfo {author} {\bibfnamefont {K.}~\bibnamefont {Blackburn}}, \bibinfo
  {author} {\bibfnamefont {C.-J.}\ \bibnamefont {Haster}}, \bibinfo {author}
  {\bibfnamefont {M.}~\bibnamefont {P\"urrer}}, \bibinfo {author}
  {\bibfnamefont {V.}~\bibnamefont {Raymond}}, \ and\ \bibinfo {author}
  {\bibfnamefont {P.}~\bibnamefont {Schmidt}},\ }\href {\doibase
  10.1103/PhysRevD.94.044031} {\bibfield  {journal} {\bibinfo  {journal} {Phys.
  Rev. D}\ }\textbf {\bibinfo {volume} {94}},\ \bibinfo {pages} {044031}
  (\bibinfo {year} {2016})},\ \Eprint {http://arxiv.org/abs/1604.08253}
  {arXiv:1604.08253 [gr-qc]} \BibitemShut {NoStop}%
\bibitem [{\citenamefont {Morisaki}\ and\ \citenamefont
  {Raymond}(2020)}]{Morisaki_2020}%
  \BibitemOpen
  \bibfield  {author} {\bibinfo {author} {\bibfnamefont {S.}~\bibnamefont
  {Morisaki}}\ and\ \bibinfo {author} {\bibfnamefont {V.}~\bibnamefont
  {Raymond}},\ }\href {\doibase 10.1103/physrevd.102.104020} {\bibfield
  {journal} {\bibinfo  {journal} {Physical Review D}\ }\textbf {\bibinfo
  {volume} {102}},\ \bibinfo {pages} {104020} (\bibinfo {year}
  {2020})}\BibitemShut {NoStop}%
\bibitem [{\citenamefont {Smith}\ \emph {et~al.}(2013)\citenamefont {Smith},
  \citenamefont {Cannon}, \citenamefont {Hanna}, \citenamefont {Keppel},\ and\
  \citenamefont {Mandel}}]{PhysRevD.87.122002}%
  \BibitemOpen
  \bibfield  {author} {\bibinfo {author} {\bibfnamefont {R.~J.~E.}\
  \bibnamefont {Smith}}, \bibinfo {author} {\bibfnamefont {K.}~\bibnamefont
  {Cannon}}, \bibinfo {author} {\bibfnamefont {C.}~\bibnamefont {Hanna}},
  \bibinfo {author} {\bibfnamefont {D.}~\bibnamefont {Keppel}}, \ and\ \bibinfo
  {author} {\bibfnamefont {I.}~\bibnamefont {Mandel}},\ }\href {\doibase
  10.1103/PhysRevD.87.122002} {\bibfield  {journal} {\bibinfo  {journal}
  {{Phys. Rev. D}}\ }\textbf {\bibinfo {volume} {87}},\ \bibinfo {pages}
  {122002} (\bibinfo {year} {2013})}\BibitemShut {NoStop}%
\bibitem [{\citenamefont {Pratten}\ \emph {et~al.}(2020)\citenamefont
  {Pratten}, \citenamefont {Husa}, \citenamefont {Garc{\'{\i} }a-Quir{\'{o}}s},
  \citenamefont {Colleoni}, \citenamefont {Ramos-Buades}, \citenamefont
  {Estell{\'{e}}s},\ and\ \citenamefont {Jaume}}]{Pratten_2020}%
  \BibitemOpen
  \bibfield  {author} {\bibinfo {author} {\bibfnamefont {G.}~\bibnamefont
  {Pratten}}, \bibinfo {author} {\bibfnamefont {S.}~\bibnamefont {Husa}},
  \bibinfo {author} {\bibfnamefont {C.}~\bibnamefont {Garc{\'{\i}
  }a-Quir{\'{o}}s}}, \bibinfo {author} {\bibfnamefont {M.}~\bibnamefont
  {Colleoni}}, \bibinfo {author} {\bibfnamefont {A.}~\bibnamefont
  {Ramos-Buades}}, \bibinfo {author} {\bibfnamefont {H.}~\bibnamefont
  {Estell{\'{e}}s}}, \ and\ \bibinfo {author} {\bibfnamefont {R.}~\bibnamefont
  {Jaume}},\ }\href {\doibase 10.1103/physrevd.102.064001} {\bibfield
  {journal} {\bibinfo  {journal} {Physical Review D}\ }\textbf {\bibinfo
  {volume} {102}},\ \bibinfo {pages} {064001} (\bibinfo {year}
  {2020})}\BibitemShut {NoStop}%
\bibitem [{\citenamefont {Hessels}\ \emph {et~al.}(2006)\citenamefont
  {Hessels}, \citenamefont {Ransom}, \citenamefont {Stairs}, \citenamefont
  {Freire}, \citenamefont {Kaspi},\ and\ \citenamefont
  {Camilo}}]{Hessels_2006}%
  \BibitemOpen
  \bibfield  {author} {\bibinfo {author} {\bibfnamefont {J.~W.~T.}\
  \bibnamefont {Hessels}}, \bibinfo {author} {\bibfnamefont {S.~M.}\
  \bibnamefont {Ransom}}, \bibinfo {author} {\bibfnamefont {I.~H.}\
  \bibnamefont {Stairs}}, \bibinfo {author} {\bibfnamefont {P.~C.~C.}\
  \bibnamefont {Freire}}, \bibinfo {author} {\bibfnamefont {V.~M.}\
  \bibnamefont {Kaspi}}, \ and\ \bibinfo {author} {\bibfnamefont
  {F.}~\bibnamefont {Camilo}},\ }\href {\doibase 10.1126/science.1123430}
  {\bibfield  {journal} {\bibinfo  {journal} {Science}\ }\textbf {\bibinfo
  {volume} {311}},\ \bibinfo {pages} {1901} (\bibinfo {year}
  {2006})}\BibitemShut {NoStop}%
\bibitem [{\citenamefont {Lo}\ and\ \citenamefont {Lin}(2011)}]{Lo_2011}%
  \BibitemOpen
  \bibfield  {author} {\bibinfo {author} {\bibfnamefont {K.-W.}\ \bibnamefont
  {Lo}}\ and\ \bibinfo {author} {\bibfnamefont {L.-M.}\ \bibnamefont {Lin}},\
  }\href {\doibase 10.1088/0004-637x/728/1/12} {\bibfield  {journal} {\bibinfo
  {journal} {The Astrophysical Journal}\ }\textbf {\bibinfo {volume} {728}},\
  \bibinfo {pages} {12} (\bibinfo {year} {2011})}\BibitemShut {NoStop}%
\end{thebibliography}%

\end{document}